\documentclass[conference]{IEEEtran}
\IEEEoverridecommandlockouts
\usepackage[ruled,linesnumbered,noend]{algorithm2e}
\usepackage{caption}% http://ctan.org/pkg/caption
\usepackage{color}
\usepackage{makecell}
\usepackage{comment}
\usepackage{multirow,diagbox}
\usepackage{soul}
\usepackage{framed}
\usepackage{subfigure}
\usepackage{graphicx}

\usepackage{array}
\usepackage{cite}
\usepackage{amsmath,amssymb,amsfonts}
\usepackage{algorithmic}
\usepackage{graphicx}
\usepackage{textcomp}
\usepackage{xcolor}

\DeclareMathOperator*{\argmin}{arg\,min}

\def\BibTeX{{\rm B\kern-.05em{\sc i\kern-.025em b}\kern-.08em
    T\kern-.1667em\lower.7ex\hbox{E}\kern-.125emX}}
\begin{document}

\title{{\normalsize In Proceedings of the 43rd IEEE Symposium on Security and Privacy (Oakland'22)}\\Universal 3-Dimensional Perturbations for Black-Box Attacks on Video Recognition Systems
}

\author{

\IEEEauthorblockA  {Shangyu Xie\textsuperscript{1}, Han Wang\textsuperscript{1}, Yu Kong\textsuperscript{2}, Yuan Hong\textsuperscript{1}}
\IEEEauthorblockA{
 \textsuperscript{1}Illinois Institute of Technology, 
 \textsuperscript{2}Rochester Institute of Technology}
Email: \{sxie14, hwang185\}@hawk.iit.edu, yu.kong@rit.edu, yuan.hong@iit.edu\\
 
% \and
% \IEEEauthorblockN{}
%  \textit{name of organization (of Aff.)}\\
% % City, Country \\
% \and
% \and
% \and
% \and
}

\maketitle

\begin{abstract}
Widely deployed deep neural network (DNN) models have been proven to be vulnerable to adversarial perturbations in many applications (e.g., image, audio and text classifications). To date, there are only a few adversarial perturbations proposed to deviate the DNN models in video recognition systems by simply injecting 2D perturbations into video frames. However, such attacks may overly perturb the videos without learning the spatio-temporal features (across temporal frames), which are commonly extracted by DNN models for video recognition. To our best knowledge, we propose the first black-box attack framework that generates universal 3-dimensional (U3D) perturbations to subvert a variety of video recognition systems. U3D has many advantages, such as (1) as the transfer-based attack, U3D can universally attack multiple DNN models for video recognition without accessing to the target DNN model; (2) the high transferability of U3D makes such universal black-box attack easy-to-launch, which can be further enhanced by integrating queries over the target model when necessary; 
(3) U3D ensures human-imperceptibility; (4) U3D can bypass the existing state-of-the-art defense schemes; (5) U3D can be efficiently generated with a few pre-learned parameters, and then immediately injected to attack \emph{real-time} DNN-based video recognition systems. We have conducted extensive experiments to evaluate U3D on multiple DNN models and three large-scale video datasets. The experimental results demonstrate its superiority and practicality.
\end{abstract}

% \begin{IEEEkeywords} DNN,
% Black-box Attack, Video Recognition
% \end{IEEEkeywords}

\section{Introduction}
\label{sec:intro}
Deep neural network (DNN) models have been extensively studied to facilitate a wide variety of intelligent video recognition systems, such as face recognition \cite{YangRZCWLH17}, action recognition \cite{FeichtenhoferPZ16} and anomaly detection \cite{SultaniCS18}. For instance, self-driving vehicles are equipped with many cameras to capture the visual information. Then, DNNs are adopted to accurately recognize road signs, detect and predict trajectories of pedestrians and vehicles, and thus make the driving decisions \cite{RauschHSLKH17, OnishiMSMO19}. Video anomaly detection systems \cite{chalapathy2019deep, SultaniCS18} integrate DNNs to monitor the activities under surveillance, and trigger alarms once anomalies (e.g., traffic accident, theft, and arson) are visually identified to advance the public safety. 

However, DNNs have been revealed to be inherently vulnerable to adversarial attacks, where attackers can add well-crafted imperceptible perturbations to the inputs to deviate the learning results. Such attacks are initially identified in the image domain \cite{SzegedyZSBEGF13,Moosavi-Dezfooli16,Moosavi-Dezfooli17, CoMML19, MirskyMSE19}, and have also attracted significant interests in other contexts, e.g., text understanding \cite{LiJDLW19,sunlog20}, and voice recognition \cite{carlini2016hidden,chen2019real,voiceccs20}. Similarly, adversarial perturbations to the DNNs in video recognition systems could potentially cause severe physical and financial damages. For instance, they may misdirect the DNN models in autonomous vehicles to inaccurately recognize objects and make detrimental decisions towards accidents. Furthermore, DNN-based anomaly detection models in video surveillance or CCTV might be deviated via the perturbations to misclassify anomalous activities to routine ones, and vice-versa \cite{SultaniCS18}.

Although the adversarial attacks on images have been well-explored, there are very limited works on attacking DNN models for videos \cite{wei2019sparse,LiNPSKRS19,abs-1911-09449 ,jiang2019black}, which need to address additional challenges, e.g., larger data sizes, a new set of DNN models for learning actions in the videos, different types of features extracted with additional temporal convolution, and different realizability. To our best knowledge, current video attacks \cite{LiNPSKRS19,wei2019sparse,abs-1911-09449,jiang2019black} adapt image perturbations in a frame-by-frame fashion to subvert DNNs for video classification, which have the following major limitations.

\begin{enumerate}
    
    \item Frame-by-frame image perturbations may overly perturb the videos (human perceptible), and also lack the temporal consistency in the perturbations. These make the attacks not robust against the state-of-the-art detection schemes (e.g., AdvIT \cite{xiao2019advit}). Adversarial examples crafted by \cite{LiNPSKRS19,jiang2019black,abs-1911-09449} can be accurately detected by AdvIT (as evaluated in our experiments).

    \item Frame-by-frame image perturbations may not be well aligned with the video frames (\emph{boundary effect} by misaligning the perturbation and video frames) \cite{LiNPSKRS19}. 
    
    \item Crafting adversarial examples for videos frame-by-frame results in heavy computation
    overheads and lacks universality. It limits the application to attack large-scale
    videos or streaming videos (e.g., CCTV surveillance).
\end{enumerate}

To address the above limitations, we propose a black-box attack framework that generates \emph{universal 3-dimensional (U3D)} perturbations to subvert a wide variety of video recognition systems. U3D has the following major advantages: (1) as a transfer-based black-box attack \cite{liu2016delving,papernot2017practical,cheng2019improving}, U3D can universally attack multiple DNN models for video recognition (each of which can be considered as the target model) without accessing to the target DNN models; (2) the high transferability of U3D makes such black-box attacks easy-to-launch, which can be further enhanced by integrating queries over the target model when necessary (validated); 
(3) U3D ensures good human-imperceptibility (validated by human survey); (4) U3D can bypass the existing state-of-the-art defense schemes (extended towards defending against U3D), including universal adversarial training (UAT) \cite{madry2018towards,shafahi2020universal}, detection schemes \cite{yin2019adversarial, xiao2019advit}, and certified schemes (e.g., PixelDP \cite{LecuyerAG0J19} and randomized smoothing \cite{cohen2019certified}); (5) U3D perturbations can be generated on-the-fly with very low computation overheads (e.g., $\sim$0.015s per frame) to attack DNN models for streaming videos. 

Specifically, in the attack design, we generate perturbations by maximally deviating features over the feature space representation of the DNNs while strictly bounding the maximum perturbations applied to the videos. We aim at generating more transferable adversarial examples (to be misclassified by multiple DNN models) by explicitly optimizing the attack performance w.r.t. layer-wise features of a video DNN model. Moreover, we integrate boundary effect mitigation and universality into the optimization for learning the U3D perturbations.

Different from traditional black-box attacks that may request intensive queries over the target DNN model, U3D perturbations can be efficiently derived independent of the target DNN model. 
Assuming that the adversary does not need to know the target DNN model under the black-box setting (and no need to query over the target model by default), our U3D attack computes the perturbation using a surrogate DNN model (any public DNN model, \emph{which can have very different model structure and parameters from the target model}). Such black-box attacks are realized via high transferability across multiple DNN models on different datasets (as validated in Section \ref{exp:transfer}). We have also shown that our U3D attack can integrate queries over target model when necessary (turning into a hybrid black-box attack \cite{usenix20-transfer}).

\begin{figure}[!tbh]
\centering
	\includegraphics[angle=0,width=0.65\linewidth]{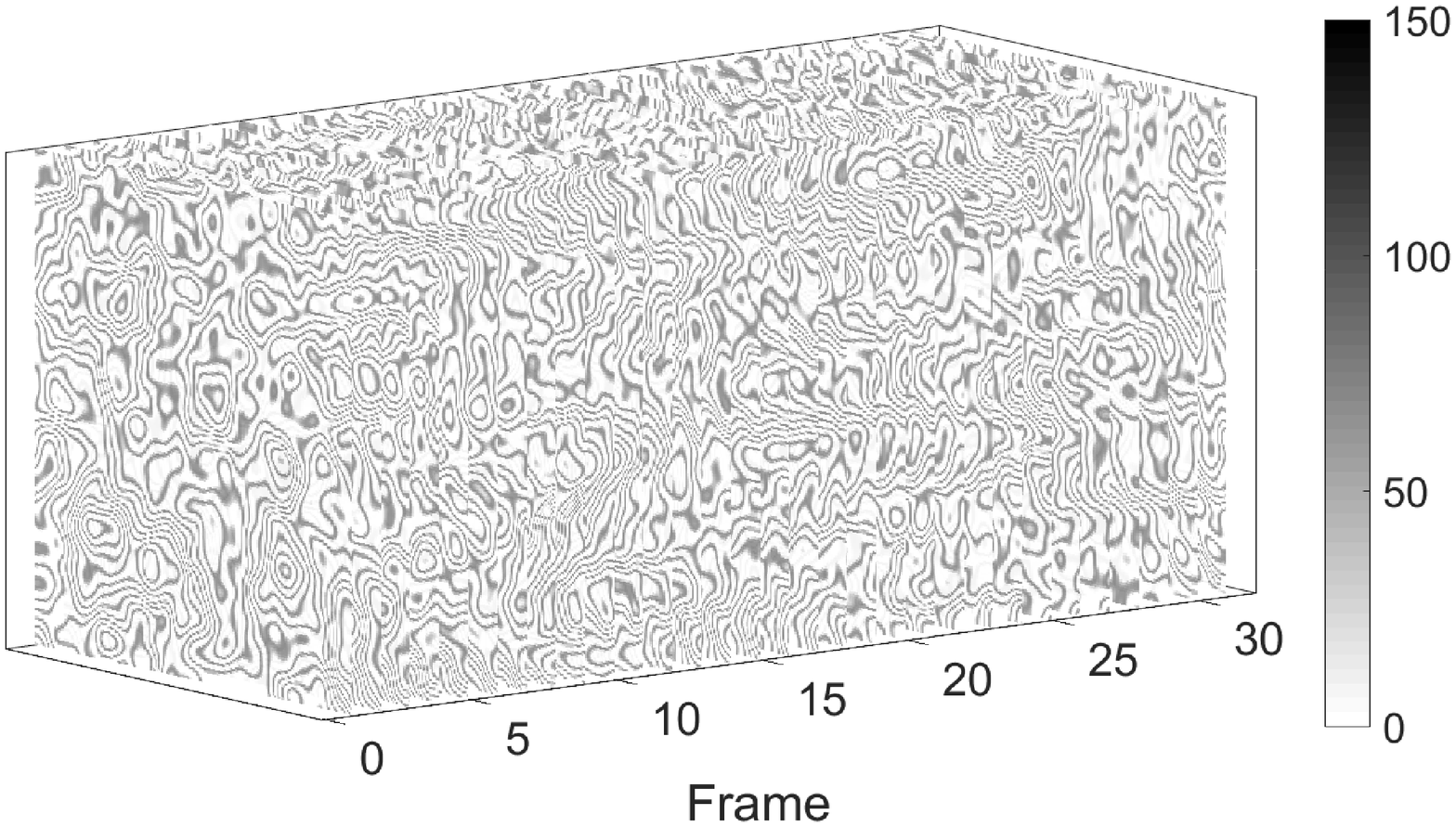}
	\caption[Optional caption for list of figures]
	{Universal 3-dimensional (U3D) perturbation}
	\label{fig:u3d}
\end{figure}

Figure \ref{fig:u3d} demonstrates an example of the U3D perturbation, which is continuously generated. Compared to the state-of-the-art universal perturbations (see Section \ref{sec:exp}), U3D achieves higher success rates with significantly less perturbations (mostly between [0,10] in grayscale [0,255]). It is also highly efficient for attacking multiple video recognition systems (e.g., classification and real-time anomaly detection). Therefore, we summarize our main contributions as below:

\begin{itemize}
  \item To our best knowledge, we propose the first black-box attack that generates 3D perturbations to universally subvert multiple DNN-based video recognition systems. 
\item We construct two different types of novel U3D perturbations optimized in the feature space representation of DNNs, which can practically attack various DNN models and the related video recognition systems (e.g., classification and anomaly detection) with high transferability.
    
    \item We conduct extensive experiments to validate the U3D attack while benchmarking with the state-of-the-art attacks (e.g., C-DUP \cite{LiNPSKRS19}, V-BAD \cite{jiang2019black} and H-Opt \cite{abs-1911-09449}). Evaluations include success rate, transferability, universality, human-imperceptibility, performance against defenses, physical realization, and efficiency. The results have shown the superiority and practicality of U3D. 
    
    \item In particular, we also evaluate the U3D against different types of state-of-the-art defense schemes. We have extensively adapted the defenses w.r.t. U3D, and studied the potential mitigation of the U3D. The high attack performance against defenses reveals the potential severity of the adversarial attack and the vulnerabilities in the DNN-based video recognition systems. Our novel U3D attack can facilitate the development of more robust and trustworthy DNN models for video recognition.

\end{itemize}

The remainder of this paper is organized as follows. We first briefly introduce the video recognition systems and define our threat model in Section \ref{sec:bg}. Section \ref{sec:models} presents the U3D design goals and attack framework. Then, we give the detailed design of the U3D attack in Section \ref{sec:classifier}. 
Section \ref{sec:exp} demonstrates the experimental results on real datasets. Section \ref{sec:disc} discusses the mitigation of the U3D attack. 
Section \ref{sec:related} reviews the related works. 
Finally, we draw conclusions in Section \ref{sec:concl}.

\section{Background}
\label{sec:bg}

\subsection{DNN-based Video Recognition Systems} \label{sec:dnn-video}
DNNs have been widely adopted for accurate video recognition in numerous real-world applications, e.g., anomaly detection \cite{SultaniCS18}, self-driving vehicles \cite{OnishiMSMO19} and smart security cameras \cite{khandare2010mobile}. There have been a series of works on designing video DNNs to improve model accuracy \cite{KarpathyTSLSF14, donahue2015long,SimonyanZ14,TranBFTP15}. For instance, Donahue et al. \cite{donahue2015long} proposed the long-term recurrent convolutional networks (LRCNs) for video recognition and description via combining convolutional layers and long-range temporal recursion. Moreover, two-stream network (TSN) \cite{SimonyanZ14} fusing static frames and optical flows was proposed for action recognition. Later, Tran et al. \cite{TranBFTP15} proposed the C3D model to significantly improve classification accuracy by focusing on spatio-temporal feature learning with 3D convolutional neural network. Recently, more networks built on spatio-temporal convolutions (e.g., I3D \cite{CarreiraZ17}) have been exhibited high performance, which greatly promoted the video recognition systems. Two example applications are demonstrated in Appendix \ref{sec:vrs}.

\subsection{Threat Model}\label{sec:threat}

\noindent\textbf{Attack Scenarios}. The U3D attack is applicable to the \emph{offline} scenario, which is identical to the attack scenario of adversarial perturbations for other types of data, e.g., images \cite{Moosavi-Dezfooli16,CoMML19,MirskyMSE19}, texts \cite{LiJDLW19}, and audio signals \cite{carlini2016hidden,chen2019real,voiceccs20}. For instance, the adversary can craft adversarial examples by adding the pre-generated U3D perturbations to static videos. Then, the perturbed videos will be misclassified to wrong labels. 

Furthermore, our U3D attack can work \emph{online} to perturb the streaming video (e.g., real-time anomaly detection in CCTV surveillance). This is also feasible since our U3D perturbations are designed to universally perturb any video at any time (from any frame in the streaming video) without the boundary effect. Thus, the U3D perturbations can be generated offline and injected into the online videos in real-time applications.

\vspace{0.05in}

\noindent\textbf{Adversary's Capabilities}. The adversary can either craft adversarial examples offline on static videos, or inject the U3D perturbations (pre-learned) into the streaming videos, similar to the attack setting in \cite{LiNPSKRS19,MirskyMSE19}. Specifically, the adversary can manipulate the systems via malware, or perform man-in-the-middle (MITM) attack to intercept and perturb the streaming videos. Furthermore, the adversary could also slightly delay the streaming video when performing injections without affecting the overall quality of the streaming video.

Note that MITM adversary is unable to perform attacks by simply replacing streaming videos with pre-recorded videos or static frames while ensuring the stealthiness of the attack, since the adversary does not know what will happen in the future \cite{LiNPSKRS19}. For instance, if the adversary wants to fool the video surveillance system in a parking lot, 
he/she may need to replace the video streams in long run (ideally all the time) to perform the attack. However, without prior knowledge on the future objects/events in the parking lot, it would be very hard to make the replaced video visually consistent with the real scenario (e.g., moving vehicles, humans, and weather). Then, the replaced video can be easily identified by the security personnel. Instead, U3D attack can be easier to be covertly realized (always human-imperceptible). The universal and boundary effect-free perturbation will be efficiently generated and continuously injected in real time (see our design goals in Section \ref{sec:design}). Thus, it can universally attack video streams in long run even if video streams may differ at different times.

We experimentally study the practicality of attack vectors (e.g., man-in-the-middle attack) in a video surveillance system \cite{costin2016security,obermaier2016analyzing,IPCamera} and implement the real-time attack based on U3D. The results show that U3D is efficient to attack real-time video recognition systems (as detailed in Section \ref{sec:real}).

\vspace{0.05in}

\noindent\textbf{Adversary's Knowledge (black-box)}. Similar to other black-box transfer-based attacks \cite{liu2016delving,papernot2017practical,cheng2019improving}, the adversary does not necessarily know the structure and parameters of the target DNN model. U3D aims to generate universal perturbations that can successfully subvert a variety of DNN models, each of which can be the potential target DNN model. By default, the adversary does not need to query the learning results (e.g., classification score or label) from the target model either. 

To successfully perform the attack, the adversary will leverage the \emph{high transferability} of U3D to deviate the target DNN models. Specifically, we assume that the adversary can utilize any public DNN model as the surrogate (e.g., C3D, I3D, LRCN and TSN) and some labeled videos (e.g., from any public data such as the HMDB51 dataset \cite{KuehneJGPS11}). Such data are not necessarily included the training data of the target DNN model. \emph{The surrogate model can be very different from the target model.} Without querying the target model, the U3D attack is even easier to realize than the conventional query-based black-box attacks \cite{chen2017zoo, brendel2017decision, pmlr-v80-ilyas18a}. 

Indeed, the U3D attack can also integrate queries over the target DNN model when necessary (see such extended attack design and evaluations in Section \ref{sec:hybrid}). Thus, the transfer-based back-box attack will turn into a hybrid black-box attack \cite{usenix20-transfer}, which integrate both query-based and transfer-based attack strategies to improve the attack performance under the black-box setting. We have experimentally validated that integrating a number of queries over the target DNN model could slightly enhance the success rates.

\section{U3D Attack Methodology}
\label{sec:models}

\begin{figure*}[!tbh]
\centering
	\includegraphics[angle=0,width=0.85\linewidth]{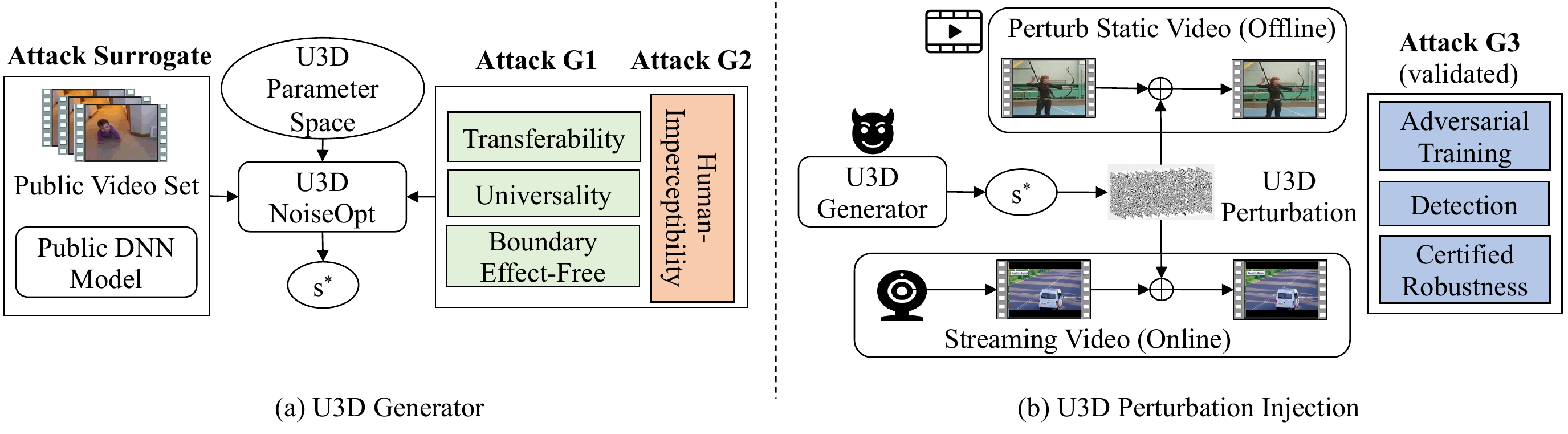}
	\caption[Optional caption for list of figures]
	{U3D attack framework (including three design goals: G1, G2 and G3). (a) $\mathtt{U3D\_Generator}$ learns the near-optimal U3D parameters $s^*$. (b) U3D perturbations can be generated on-the-fly with $s^*$ to perturb both static and streaming videos. 
	}\vspace{-0.15in}
	\label{fig.framework}
\end{figure*}

\subsection{U3D Attack Design Goals}\label{sec:design}

The goals in our U3D attack design include: 

\begin{itemize}
    \item G1: The attack should achieve high performance on the video recognition systems under the black-box setting. 
    \item G2: The adversarial perturbations should be very small to obtain good human-imperceptibility. 
    \item G3: The adversarial examples are robust against existing defense schemes (cannot be easily detected or mitigated). 
\end{itemize}

\noindent\textbf{G1: High Attack Performance}. To launch the U3D attack, the following properties are desired: (1) transferable on a wide variety of DNN models for video recognition; (2) universal on a large number of videos; (3) boundary effect-free. 

Different from increasing the magnitude of the perturbations for transferability \cite{SzegedyZSBEGF13,chen2019real,carlini2017towards}, we formulate an optimization problem with a surrogate DNN model (which can be any public DNN model) in an interpretable fashion. The objective is to maximize the distance between the clean video and perturbed video in the feature space representation (Section \ref{ssub:transfer}). First, the change of feature space representations via perturbations (especially the deep spatio-temporal feature space for videos) will non-trivially impact the classification results. This will increase the success rates of the attack. Second, the explicit attack in the feature space could craft more transferable adversarial examples since the intermediate layer features of DNNs have shown to be transferable \cite{yosinski2014transferable}. Experimental results in Section \ref{sec:exp} have demonstrated high cross-model transferability for feature space perturbations. 

Moreover, the adversary does not have prior knowledge on the video (especially the streaming video), then the 3D perturbations should universally attack \cite{Moosavi-Dezfooli17} a large number of videos (ideally, any video). We construct U3D perturbations from a relatively small set of videos to fool the target DNN model on arbitrary input videos with high success rates. 

With temporal dimensions on multiple frames, the video perturbations should address the potential misalignment with the input video (boundary effect \cite{LiNPSKRS19}), which can degrade the attack performance, especially in long run. While launching attacks, the perturbation should be effectively injected at any time in the video. To address the misalignment, we employ a transformation function to convert the perturbation temporally, and then optimize the attack on all temporal transformations (see Section \ref{ssub:boundary}), which enable the U3D perturbations to be injected at random times without the boundary effect.

\vspace{0.05in}

\noindent\textbf{G2: Human-Imperceptibility}. 
We add a bound on the U3D perturbations with $\ell_\infty$-norm, which strictly restricts the pixel deviations. Later, we use MSE metrics to quantify the perturbations in the experiments. Moreover, we conduct surveys among humans to illustrate the imperceptibility of U3D.

\vspace{0.05in}

\noindent\textbf{G3: Robustness against Defenses}. To show the robustness of our U3D attack, we implement attacks on the video recognition models equipped with defense schemes (\emph{G3 is not directly designed but ensured with post-validation}). There are two rules of thumb for evaluating attacks: (1) we should carefully utilize current effective defenses to explicitly defend against the newly proposed attack, e.g., developing adaptive schemes which uncover the potential weaknesses of the attack; (2) the defenses should be in white-box setting, i.e., the defender should be aware of the attack, including the adversary's knowledge and strategy. The rules of thumb also work for evaluating newly proposed defenses vice versa \cite{tramer2020adaptive,athalye2018robustness,carlini2017magnet}.

Specifically, we adapt three types of major defense schemes: (1) adversarial training \cite{madry2018towards, shafahi2020universal}; (2) detection  \cite{xiao2019advit,yin2019adversarial}; (3) certified robustness \cite{LecuyerAG0J19,LeCunMBCF05}. We redesign the defense schemes to defend against universal perturbations or U3D per the rules of thumb. For example, based on the adversarial training (AT) \cite{madry2018towards,shafahi2020universal}, we design the U3D-AT, which utilizes the capability of AT to defend against the best U3D (iteratively updating U3D perturbations). See details in Section \ref{exp:defend}. 

\subsection{U3D Attack Overview}
We now overview the U3D attack in Figure \ref{fig.framework}. We first formulate the U3D perturbation generation ($\mathtt{U3D\_{Generator}}$) by synthesizing the procedural noise \cite{perlin1,LagaeLCDDELPZ10} (which can be efficiently generated with low-frequency patterns) with the U3D parameters $s$ (see Section \ref{subsec:u3d_formu}). Meanwhile, the attack goals of U3D are optimized: transferability, universality, and boundary effect-free (see Section \ref{sec:cali_u3d}). Then, we apply the particle swarm optimization (PSO) to solve the problem to derive the near-optimal parameters $s^*$ for generating U3D perturbations (see Section \ref{u3d:alg}). Finally, U3D perturbations can be generated on-the-fly to be injected into the videos in the attack scenarios (either static or streaming videos). 

\subsection{U3D Attack Formulation}
\label{sec:u3d}

The DNN model can be modeled as a function $f(\cdot)$ that infers the video $v$ with a label (e.g., the label with the top-1 probability). The attack crafts a video adversarial example $v^\prime$ by injecting the perturbation $\xi$ into the original video $v$: $v^\prime=v+\xi$, where the output label of $v^\prime$ by the DNN model $f(\cdot)$ would be $f(v') \neq f(v)$ (as a universal attack). 

To pursue human-imperceptible perturbations, $\ell_{\infty}$-norm is adapted to bound the distance between the original and perturbed videos (w.r.t. U3D perturbation $\xi$) with a pre-specified small value $\epsilon$: $||v^{\prime}-v||_{\infty}=\max_i|\xi| \leq \epsilon$. Then, we formulate an optimization problem to generate U3D perturbations: $\argmin_\xi: \Gamma(v+\xi),s.t.~||\xi||_{\infty}\leq \epsilon$, where $\Gamma$ is a loss metric function, e.g., a distance or cross-entropy metric. In Section \ref{sec:cali_u3d}, we align the objective function with the attack goals in the optimization for the U3D design.

\section{Attack Design}
\label{sec:classifier}

\subsection{U3D Perturbation Formalization}
\label{subsec:u3d_formu}

``Procedural noise'' \cite{perlin1, perlin2,LagaeLCDDELPZ10,LagaeLDD09} refers to the algorithmically generated noise with a predefined function, which can be added to enrich visual details (e.g., texture, and shading) in computer graphics. It can be directly computed with only a few parameters, and has no noticeable direction artifacts \cite{perlin1, LagaeLCDDELPZ10, CoMML19}. These properties make it \emph{potentially fit for inexpensively computing adversarial perturbations}. 
While constructing U3D perturbations, we utilize two types of common procedural noises: (1) ``Perlin noise'' \cite{perlin1, perlin2} (a lattice gradient noise) due to its ease of use, popularity and simplicity; (2) ``Gabor noise'' \cite{LagaeLDD09} (a convolutional sparse noise) with good sparsity and accurate spectral control. We propose two types of U3D perturbations, ``U3D$_p$'' and ``U3D$_g$'', both of which universally perturb videos to subvert the DNN models.

We first formally define the U3D noise function. Denote $\mathcal{N}(x, y, t; S)$ as the U3D noise function, where $(x, y, t)$ represents the 3D coordinates of each pixel in the video, and $S$ is the parameter set for noise generation. 

\subsubsection{U3D$_p$ Noise}

Perlin noise \cite{perlin1, perlin2} originally works as an image modeling primitive to produce natural-looking textures in realistic computer generated imagery.

Specifically, we denote every pixel in a video by its 3D coordinates $(x,y,t)$ where $(x,y)$ are the coordinates in frame $t$, and denote the Perlin noise value of the pixel $(x, y, t)$ as $p(x, y, t)$. To model the change of visual perturbations, we define three new parameters of wavelength $\lambda_x$, $\lambda_y$, $\lambda_t$ to determine the octaves along the three dimensions x-axis, y-axis, and frame $t$, respectively, and define the number of octaves as $\Lambda$. The newly updated noise is computed as the sum of all the corresponding octaves for 3D coordinates:

\begin{equation}
\small
    \mathcal{N}(x,y,t) = \sum_{\ell=0}^{\Lambda} p(x \cdot \frac{2^{\ell}}{\lambda_{x}}, y \cdot \frac{2^{\ell}}{\lambda_{y}}, t \cdot \frac{2^{\ell}}{\lambda_{t}})
    \label{eq:perlin}
\end{equation}

Moreover, we compose the noise function with a color map function \cite{szafir2017modeling} to generate distinct visual perturbations in the video. Then, the noise of pixel $(x,y,t)$ can be derived as:

\begin{equation}
\small
    \mathcal{N}_p(x,y,t) = cmap(\mathcal{N}(x,y,t), \phi) 
    \label{eq:color}
\end{equation}

where $cmap(p, \phi) = \sin(p\cdot2 \pi \phi)$ is a sine color map function, which ensures \emph{the bound of noise value with the circular property}. $\phi$ indicates the period of the sine function, and the visualization of perturbations can be tuned with $\phi$. 

\vspace{0.05in}

\noindent\textbf{U3D$_p$ Parameters}. Combining Equation \ref{eq:perlin} and \ref{eq:color}, we denote the corresponding parameter set as $S_p$ for U3D$_p$ noise:  

\begin{equation}
\small
    S_p=\{\lambda_x, \lambda_y, \lambda_t, \Lambda, \phi \}
    \label{eq:pt}
\end{equation} 

\subsubsection{U3D$_g$ Noise}

Gabor noise \cite{LagaeLDD09, LagaeLCDDELPZ10} is a type of sparse convolution noise that obtains a better spectral control via a Gabor kernel, a multiplication of circular Gaussian envelope and a harmonic function \cite{gabor1946theory}. We construct U3D$_g$ noise by first extending the 2D Gabor kernel to 3D Gabor kernel (adding the temporal dimension $t$):

\begin{equation}
\small
    g(x,y,t) = Ke^{-\pi\sigma^2(x^2+y^2+t^2)}\cos{\left[2\pi F(x^{\prime}+y^{\prime}+t^{\prime})\right]} \label{eq:kernel}
\end{equation}

where $x^{\prime}=x\sin\theta\cos{\omega},y^{\prime}=y\sin\theta\sin{\omega}, t^{\prime}=t\cos\theta$; $K$ and $\sigma$ are the magnitude and width of the Gaussian envelope; $F$ and ($\theta$, $\omega$) are the magnitude and orientation angles of the frequency in the harmonic function. Then, we derive the noise $\mathcal{N}(x,y,t)$ with the sparse convolution and 3D Gabor kernel: 

\vspace{-0.1in}

\begin{equation}\small
    \mathcal{N}(x, y, t) = \sum_k g(x-x_k, y-y_k, t)
\end{equation}

\vspace{-0.02in}

where the point set $\{\forall (x_k, y_k, t)\}$ are a set of sampled pixel points in the same frame $t$ with Poisson distribution. Furthermore, to model the isotropy of the Gabor noise \cite{LagaeLCDDELPZ10}, we realize the two frequency orientations $(\theta, \omega)$ as random variables $(\theta_i, \omega_i)$ uniformly distributed in $[0, 2\pi]$. Then, the updated U3D$_g$ noise is given as below:

\vspace{-0.1in}

\begin{equation}\small
    \mathcal{N}_g(x, y, t) = \sum_{i} \mathcal{N}(x, y, t; (\theta_i, \omega_i)) \label{eq:gabor}
\end{equation}

\noindent \textbf{U3D$_g$ Parameters}. Similar to U3D$_p$, we denote the following parameter set as $S_g$ for U3D$_g$ with Equation \ref{eq:kernel} and \ref{eq:gabor}: 
\begin{equation}\small
    S_g =\{K, \sigma, F\}
    \label{eq:gt}
\end{equation} 

We synthesize the procedural noise to construct the U3D perturbations, whose low-frequency patterns and low computational overhead can greatly advance the attacks. Formally, given the U3D noise function $\mathcal{N}$ and the parameters $s$, the generated U3D perturbation $\xi$ of length $T$ will be:

\vspace{-0.1in}

\begin{equation}\small
    \xi=\{\mathcal{N}(t;s)|t\in[0,T-1]\} \label{eq:x}
\end{equation}

If $T$ is less than the video length, $\xi$ will be circular. Note that $T$ works as a pre-specified parameter. For simplification, we use $\xi=\mathcal{N}(T;s)$ to represent Equation \ref{eq:x}. Next, we will present how to calibrate U3D perturbation to achieve the design goals.

\subsection{Calibrating U3D Perturbations}\label{sec:cali_u3d}

\subsubsection{Improving Transferability in Feature Space}\label{ssub:transfer}

U3D aims to deviate the intermediate layer's features, which could improve the transferability of the attacks. Large distance between the original and perturbed videos' features at intermediate layers of the DNN model can result in relatively high deviations in the final results. This will increase the probabilities on false learning by the unknown target DNN model and videos.

Specifically, we formally define $f_L(\cdot, d)$ as the truncated DNN model function, which outputs the intermediate feature of the input video at layer $L_d, d\in[1,M]$ of the DNN model $f(\cdot)$, $M$ is the number of DNN layers. Then, $f_L(v, d), f_L(v^{\prime}, d)$ are denoted as the intermediate features of the original video $v$ and perturbed video $v^{\prime}$, respectively. Thus, we have the $\ell_2$-norm distance between the feature representations of the original video $v$ and perturbed video $v^{\prime}=v+\xi$ at layer $d$ of the DNN as: 

\vspace{-0.07in}

\begin{equation}\small
   \mathcal{D}(v,v';d)= ||P(f_L(v, d))-P(f_L(v', d))||_2
\end{equation}

\vspace{-0.02in}

where $P(z)=sign(z)\odot |z|^\alpha$ is a power normalization function $\alpha\in[0, 1]$ and $\odot$ is the element-wise product \cite{perronnin2010improving}. 

Then, we maximize the distance $\mathcal{D}(v,v^{\prime};d)$ between the original and perturb videos over all the intermediate feature space as our attack objective function:

\vspace{-0.07in}

\begin{equation}\small
\begin{aligned}
    \max_\mathcal{\xi}: \sum_{d\in [1,M]}\mathcal{D}(v,v+\xi;d)
\end{aligned}    
\label{eq:u3d_dista}
\end{equation}

\vspace{-0.02in}

\subsubsection{Mitigating Boundary Effect} \label{ssub:boundary}

Recall that the boundary effect may potentially degrade the attack performance due to the misalignment between the adversarial perturbation and the input video. 
To tackle such issue, we introduce a temporal transformation function $\mathtt{Trans}(\cdot)$ for the U3D perturbation with a shifting variable denoted as $\tau$. Specifically, given a U3D perturbation $\xi$ of length $T$, then $\mathtt{Trans}(\xi; \tau)$ represents the U3D perturbation $\xi$ temporally shifted by $\tau\in[0, T-1]$. Then, we maximize the expectation of the feature distances with all the $T$ possible temporal shift transformation $\tau\in U[0,T-1]$ for U3D perturbation $\xi$ ($U$ denotes the uniform distribution): 

\vspace{-0.07in}

\begin{equation}\small
\begin{aligned}
    \max_\mathcal{\xi}:& \mathop{\mathbb{E}}_{\tau\sim U[0,T-1]}[\sum_{d\in M}\mathcal{D}(v,v+\mathtt{Trans}(\xi, \tau);d)] 
\end{aligned}    
\label{eq:u3d_shift}
\end{equation}

\vspace{-0.02in}

To achieve such objective, we can consider all the possible transformed U3D perturbation (the size of transformation will be $T$) uniformly shifted with $\tau\in[0, T-1]$ (step $1$ frame by frame in the video). $\tau$ will be sampled in the corresponding algorithm. Then, our U3D attack can learn a generic adversarial perturbation without the boundary effect, which can be injected into the streaming video anytime.

\subsubsection{Improving Universality with Public Videos}\label{ssub:univer} 

Another goal is to find a universal perturbation learned from a relatively small set of videos, which can effectively perturb the unseen videos for misclassification. Denoting a set of public videos as $V$, the optimization integrates the universality maximization on videos in $V$ (and $\ell_\infty$-norm bound) as below:

\vspace{-0.1in}

\begin{equation}\small
\begin{aligned}
    \max_\mathcal{\xi}:& \mathop{\mathbb{E}}_{v\sim V, \tau\sim U[0,T-1]}[\sum_{d\in M}\mathcal{D}(v,v+\mathtt{Trans}(\xi, \tau);d)] \\s.t.&~~\xi=\mathcal{N}(T; s),~ ||\xi||_{\infty}\leq\epsilon
\end{aligned}    
\label{eq:u3d_opt_total}
\end{equation}

\subsection{Optimizing and Generating U3D Perturbations}
\label{u3d:alg}

Since the U3D perturbation $\xi$ can be efficiently generated if the U3D parameter set $\mathcal{S}$ is pre-computed, Equation \ref{eq:u3d_opt_total} will optimize the attack w.r.t. $\mathcal{S}$. To search the optimal U3D parameter set $\mathcal{S}$), we solve it with the Particle Swarm Optimization (PSO) method \cite{eberhart1995new}. Specifically, the parameter values in $\mathcal{S}$ are viewed as the particles' positions, and the set of parameter ranges can be constructed as the search space. Then, the objective function (Equation \ref{eq:u3d_opt_total}) is the fitness function $\mathcal{A}(f, V, \mathcal{N}(T;\vec{s}))$, where $\vec{s}$ is the current position for U3D parameter set $S$ in the iterations. 

In the initialization phase, $m$ points will be randomly selected from the searching space for $\mathcal{S}$ while satisfying $\ell_\infty$-norm bound. Then, in the iterations, every particle will iteratively update its position by evaluating the personal and group best location (determined by the output of the fitness function). Notice that, before fed into the fitness function, the algorithm validates if the U3D perturbation $\xi$ generated by the parameter set $\vec{s_i}^{k+1}$ satisfies $\ell_\infty$-norm bound $\epsilon$ or not. Finally, we can get the near-optimal parameter set $s^*$. Then, we generate the U3D perturbation $\xi=\mathcal{N}(T;s^*)$. The computation of fitness function $\mathcal{A}(f, V, \mathcal{N}(T;\vec{s}))$ and PSO optimization process are detailed in Algorithm \ref{alg:oracle} and \ref{alg:swarm}, respectively (Appendix \ref{sec:pso}).

We have evaluated PSO by benchmarking with genetic algorithms \cite{goldberg1988genetic}, simulated annealing \cite{van1987simulated}, and Tabu search \cite{glover1989tabu}. PSO slightly outperforms them for U3D optimization (experimental setting and results are detailed in Appendix \ref{app:pso_performance}).

\section{Experiments}
\label{sec:exp}

\subsection{Experimental Setup}
\label{sec:setup}

\noindent\textbf{Datasets}. We use three widely used datasets for video recognition to validate the proposed U3D attack. 
\begin{itemize}
\item HMDB51 \cite{KuehneJGPS11} dataset includes 6,766 video clips (30 fps) in 51 different actions, e.g., fencing, climb and golf. 
\item UCF101 \cite{abs-1212-0402} dataset includes 13,320 video clips (25 fps) in 101 different actions, e.g., archery, and punch. 
\item UCF Crime \cite{SultaniCS18} dataset includes 1,900 long surveillance videos (30 fps) collected from Youtube and Liveleak, in 13 anomalies, e.g., accident, explosion, and shooting.
\end{itemize} 

The HMDB51 and UCF101 datasets are used for video classification, and the UCF Crime dataset for anomaly detection.

\vspace{0.05in}

\noindent\textbf{Target DNN Models}. We evaluate the U3D attack on two common DNN models for video recognition: (1) C3D model \cite{TranBFTP15}; (2) I3D model \cite{SultaniCS18}. We also implement two video recognition techniques based on both C3D and I3D:  (1) video classification \cite{TranBFTP15}; (2) video anomaly detection \cite{SultaniCS18} identifying anomalies by scoring the video segments in sequence. 

Note that we choose C3D and I3D as the main evaluation models to show the attack performance due to the popularity and practicality in the video recognition systems (as depicted in Section \ref{sec:dnn-video}). To fully evaluate the \emph{transferability} of the U3D attack, we choose three more video classification models, including LRCN \cite{donahue2015long}, DN \cite{KarpathyTSLSF14} and TSN \cite{SimonyanZ14}, and evaluate the U3D attack across five different DNN models. We summarize the differences of such five models in Appendix \ref{app:difference}.

\vspace{0.05in}

\noindent\textbf{Benchmarks}. We use the following baseline adversarial perturbations: (1) Gaussian Noise: $\xi_g\sim\mathcal{N}(0, \sigma^2)$ and $\sigma$=0.01; (2) Uniform Noise: uniformly sampled noise $\xi_u\sim[-\epsilon$, $\epsilon]$; (3) Random U3D: applying U3D without calibration by randomly choosing parameters. For the above three methods, we repeat each experiment 10 times, and return the average value; (4) The state-of-the-art video attacks, C-DUP \cite{LiNPSKRS19} (as a \emph{white-box} universal attack), V-BAD \cite{jiang2019black} and H-Opt \cite{abs-1911-09449} (both as \emph{non-universal} black-box attacks). 

Since V-BAD \cite{jiang2019black} and H-Opt \cite{abs-1911-09449} are non-universal, they might be incomparable with U3D and C-DUP on attacking a specific target (though their success rates are claimed to be high in such cases). It might also be unfair to compare U3D and C-DUP with V-BAD and H-Opt on transferability since the latter two are not designed towards that goal.

\subsection{Attack Performance}\label{exp:effect}

We first evaluate U3D$_p$ and U3D$_g$ generated with a surrogate C3D model to attack unknown target models on different datasets. Specifically, we randomly select 500 videos from the HMDB51 dataset (retaining a similar distribution for classes as the full dataset) as the public video set ($V$), and consider the full UCF101 and UCF Crime datasets as the target set. We set $\epsilon=8, T=16$ and report the attack results on the target set. Note that the MSE of all the U3D perturbations are below 20 (very minor distortion out of 255$^2$ in the scale). 

\begin{table}[!h]
 \centering\small
 \caption{U3D vs. benchmarks (success rates; C3D/HMDB51 as surrogate; C3D/I3D and UCF101/UCF Crime as target).}
  \begin{tabular}{c|cc|cc}
  \hline
  
\multirow{2}{*}{\diagbox[width=6em]{Noise}{Model}} & \multicolumn{2}{c|}{C3D}&\multicolumn{2}{c}{I3D}\\\cline{2-3}\cline{4-5}& UCF101 &UCF Crime & UCF101 &UCF Crime\\
\hline
 
{Gaussian}& 10.2\% &  15.3\% &   9.1\% & 12.6\% \\ 
Uniform& 5.3\% & 9.1\% &  1.7\% &2.4\% \\
 Rnd. U3D &43.2\%& 52.6\% &40.3\% & 51.8\%\\
 {C-DUP\cite{LiNPSKRS19}} &80.2\% &83.6\%&54.4\%&45.8\% \\
\hline
 {U3D$_p$}  &82.6\%&92.1\%&80.4\%&87.1\%\\
 {U3D$_g$} &85.4\% &93.4\% &82.9\%&90.2\%  \\\hline
 
  \end{tabular}\vspace{-0.05in}
  \label{tab:attack_result}
\end{table}

Table \ref{tab:attack_result} lists the results for applying U3D$_p$ and U3D$_g$ to attack unknown models and videos (in different datasets). The U3D perturbations are injected into both UCF101 (for \emph{video classification}) and UCF Crime (for \emph{anomaly detection}) datasets, which are then inferred by both C3D and I3D. Such black-box attacks are realized by the \emph{transferability} and \emph{universality} of U3D (which will be thoroughly evaluated in Section \ref{exp:transfer}). 
Table \ref{tab:attack_result} also includes the attack performance of Gaussian, Uniform, Random, and C-DUP \cite{LiNPSKRS19} (see the setting in Section \ref{sec:setup}). For both U3D and benchmarks, we apply the perturbations to full UCF101 and UCF Crime datasets.

Both U3D$_p$ and U3D$_g$ achieve high success rates on the two DNNs. For C3D, U3D$_p$ achieves 82.6\% on the UCF101 dataset (video classification) and 92.1\% on the UCF Crime (anomaly detection) while U3D$_g$ obtains a slightly higher success rate, i.e., 85.4\% on the UCF101, and 93.4\% on the UCF Crime. This can also show that our U3D perturbations effectively attack to other different DNN models on different datasets, e.g., HMDB51 and C3D $\to$ UCF Crime and I3D. 

However, the benchmarks cannot achieve satisfactory attack performance. Injecting random noise (Gaussian and Uniform) to videos can only give 2.4\%-15.3\% success rates in all the experiments. Random U3D (random parameters without optimization) performs better but still not satisfactory (35.7\%-52.6\%). C-DUP \cite{LiNPSKRS19} returns worse success rates on both C3D and I3D, even in the white-box setting. Since it is designed for attacking C3D, its performance on I3D is even worse (54.4\% on UCF101 and 45.8\% on UCF Crime, low transferability). 

Finally, both U3D$_p$ and U3D$_g$ can perform very well (90\%+) on anomaly detection regardless of the target models. We observe that anomaly detection is more vulnerable than video classification under the same perturbation, e.g., U3D$_p$ (92.1\%$>$82.6\%).  The possible reason is that such DNN models have an extra computing model to output anomaly scores, which may make it more sensitive to perturbations. 

\subsection{Transferability and Universality}\label{exp:transfer}

\noindent\textbf{Transferability}. \emph{Transferability} refers to the perturbations designed for one classifier can also attack other classifiers (cross-model) \cite{SzegedyZSBEGF13}. To study the transferability, we first define the \emph{transfer rate} (TR) as the percent of the adversarial examples which deviates one model $f_{srg}$ (e.g., a public DNN model as the surrogate model) and also deviate the target model $f_{tar}$ (black-box). We denote $f_{srg}\to f_{tar}$ as the transferability of the attack from surrogate model to target model.

\vspace{-0.05in}

\begin{table}[htbp]
 \centering\small
 \caption{Transferability: transfer rate (TR) on UCF101 from surrogate model $f_{srg}$ to target model $f_{tar}$. See similar results on HMDB51 and UCF Crime in Appendix \ref{app:add}.}
 \begin{tabular}{c|c|ccccc}
  \hline
Noise &\diagbox[width=4em]{$f_{srg}$}{$f_{tar}$} &C3D&I3D& DN & LRCN  & TSN\\  
  \hline
   \multirow{5}{*}{U3D$_p$}&C3D&--&93.4\%  & 92.7\% &85.0\% &87.2\%\\
   &I3D&89.7\%  &--& 96.3\% &88.7\% &85.0\%\\
   &DN&84.0\%  &83.2\% & --& 85.5\% &83.4\%\\
   &LRCN&85.8\%  &87.2\%&92.4\%&-- &86.1\%\\
   &TSN &85.5\% & 82.5\%  & 89.3\% &87.5\% &--\\
   \hline
  \multirow{5}{*}{U3D$_g$}&C3D&--&87.0\% &  93.2\%  & 86.3\% & 85.3\%\\
   &I3D&88.2\%  &  --&  97.4\%  & 85.2\% & 86.0\%\\
   &DN&82.6\%  &81.4\% & --& 83.7\% &85.6\%\\
   &LRCN&81.2\%  &83.4\%&88.6\%&-- &84.5\%\\
   &TSN &86.2\% & 83.6\%  & 90.2\% &86.4\% &--\\
  \hline

  \end{tabular}
  \label{tab:cross-model}\vspace{-0.05in}
\end{table}

To evaluate the transferability, we choose C3D, I3D and other three more video classification models as surrogate/target models: DN \cite{KarpathyTSLSF14}, LRCN \cite{donahue2015long}, and TSN \cite{SimonyanZ14}, all of which are already fine-tuned on the UCF101 dataset. Then, we compute U3D$_p$ and U3D$_g$ ($\epsilon=8$) with the surrogate model (as surrogate) and apply the U3D perturbations to craft adversarial examples on the UCF101 dataset, which are fed into the target models. We generate the U3D perturbations with 10\% of the UCF101 dataset (randomly picked for each class), and select all the adversarial examples (crafted on the 90\% of videos for target dataset) which can successfully fool the surrogate models. Then, we examine the percent of such adversarial examples that can fool the target model. 

Table \ref{tab:cross-model} presents the transfer rates of both U3D$_p$ and U3D$_g$. We can observe that all the attack can achieve high transfer rates (over 80\%). This shows that our U3D perturbations achieve good transferability across these models. For example, U3D$_p$ can obtain 92.7\% transfer rate for C3D$\to$DN and 92.4\% for LRCN$\to$DN. We repeat the same set of experiments on the HMDB51 and UCF Crime datasets (results are given in Table \ref{tab:cross-model1} and \ref{tab:cross-model2} in Appendix \ref{app:add}). High cross-model transferability for U3D can also be observed from such experiments. 

\vspace{0.05in}

\noindent\textbf{Universality}. \emph{Universality} refers to the perturbations learnt from one dataset can perturb many different datasets (cross-data) to fool the DNN models \cite{Moosavi-Dezfooli17}. To evaluate the universality of U3D, we first randomly pick 500 videos as the surrogate video set (denoted as $X$) from each of the three datasets, HMDB51, UCF101 and UCF Crime, respectively (retaining a similar class distribution as the full datasets). Then, we compute the U3D perturbations ($\epsilon=8$) on $X$ with the C3D model and evaluate the attack success rates on all the three datasets (as the target dataset $Y$). For the same dataset case (intra-dataset attack), the target set $Y$ excludes $X$ to evaluate the universality. All the results are listed in Table \ref{tab:cross-dataset}. 

We can observe that the U3D achieve 80\%+ success rates for all the cases ($X\to Y$ within the same dataset or across different datasets). The diagonal results are higher than other cases, which shows that the U3D can perform well among the unseen videos in the same dataset. Moreover, for the same U3D perturbation, e.g., U3D$_g$, the success rate of UCF Crime$\to$UCF101 is lower than that of HMDB51$\to$UCF101 (81.6\%$<$85.4\%). Similar results are also observed from UCF Crime$\to$HMDB51 and UCF101$\to$HMDB51 (82.2\%$<$85.0\%). Since HMDB51 and UCF101 consist of human action videos while UCF Crime includes surveillance videos for anomaly detection, U3D perturbations learned on UCF Crime will exhibit less universality to HMDB51 and UCF101. Thus, selecting different surrogate videos can slightly help tune the attack performance on different target models and videos.

\begin{table}[htbp]
 \centering\small
 \caption{Universality (success rate (SR); surrogate C3D). See similar results for surrogate I3D in Appendix \ref{app:add}.}
  \begin{tabular}{c|c|ccccc}
  \hline
  
Noise &\diagbox[width=6em]{$X$}{$Y$} & HMDB51 & UCF101 &  UCF Crime \\
  \hline
   &HMDB51 & 87.3\% & 82.6\% & 92.1\%\\
    U3D$_p$ &UCF101 & 84.2\% & 88.4\% & 91.5\%\\
     &UCF Crime & 80.1\% & 82.4\% & 96.0\%\\\hline
   &HMDB51& 88.7\% & 85.4\% & 93.4\%\\
     U3D$_g$ &UCF101 & 85.0\% & 86.2\% & 90.2\%\\
     &UCF Crime & 82.2\% & 81.6\% & 95.3\%\\
    
  \hline
  
  \end{tabular}\vspace{-0.05in}
  \label{tab:cross-dataset}
\end{table}

We repeat the same set of experiments for I3D as surrogate (see similar high universality in Table \ref{tab:cross-dataset1} in Appendix \ref{app:add}). Note that C-DUP \cite{LiNPSKRS19} (as a \emph{white-box} attack only on C3D) has low transferability (in Table \ref{tab:attack_result}), and V-BAD \cite{jiang2019black} and H-Opt \cite{abs-1911-09449} (both as \emph{non-universal} attacks) have low transferability.

\subsection{Hybrid Black-Box Attack with Queries over Target Model}
\label{sec:hybrid}
U3D is designed to universally attack different target models, and it has shown high transferability. If the attack performance is not very satisfactory for a new target model (though not found in our extensive experiments), we can extend the U3D to a hybrid black-box attack \cite{usenix20-transfer} by integrating queries over the target model $g(\cdot)$. Note that this still maintains attack universality on different target models. Thus, given the surrogate model $f(\cdot)$ (including a small set of public videos) and the target model $g(\cdot)$ available for queries, we can update the optimization for U3D by integrating the queries using input videos $v_1, \dots, v_n$ (perturbed by $\xi$ before querying):

\vspace{-0.1in}

\begin{equation}
\small
\begin{aligned}
\max_{\xi}: & \mathop{\mathbb{E}}_{v\sim V, \tau\sim U[0,T-1]}[\sum_{d\in M}\mathcal{D}(v,v+\mathtt{Trans}(\xi, \tau);d)]\\
&~~~~~~+\omega\cdot\mathcal{Q}(g, v_1,\dots, v_n, \xi) \\ 
s.t.&~~\xi=\mathcal{N}(T;s), ~\|\xi\|_\infty \leq \epsilon
\label{eq:opt}
\end{aligned}
\end{equation}

where the query oracle $\mathcal{Q}(\cdot)$ first derives the final classification results of the perturbed videos $\{v_1+\xi,\dots\, v_n+\xi\}$ by the target model $g(\cdot)$, and then returns the success rate for such videos. $\omega$ is hyperparameter for weighing the transferability of the surrogate model and queries (success rate) over the target model. Note that the adversary only needs to query the final classification (limited information) instead of the specific probability scores or logits information. 

This optimization will search the U3D perturbations which can successfully fool the target model $g(\cdot)$ by both transferability and queries (hybrid). Similarly, after revising the fitness function with the new objective (Equation \ref{eq:opt}), we can apply PSO to compute the optimal U3D parameters.

\vspace{-0.15in}

\begin{figure}[!h]
	\centering
		\subfigure[ U3D$_p$]{
		\includegraphics[angle=0, width=0.495\linewidth]{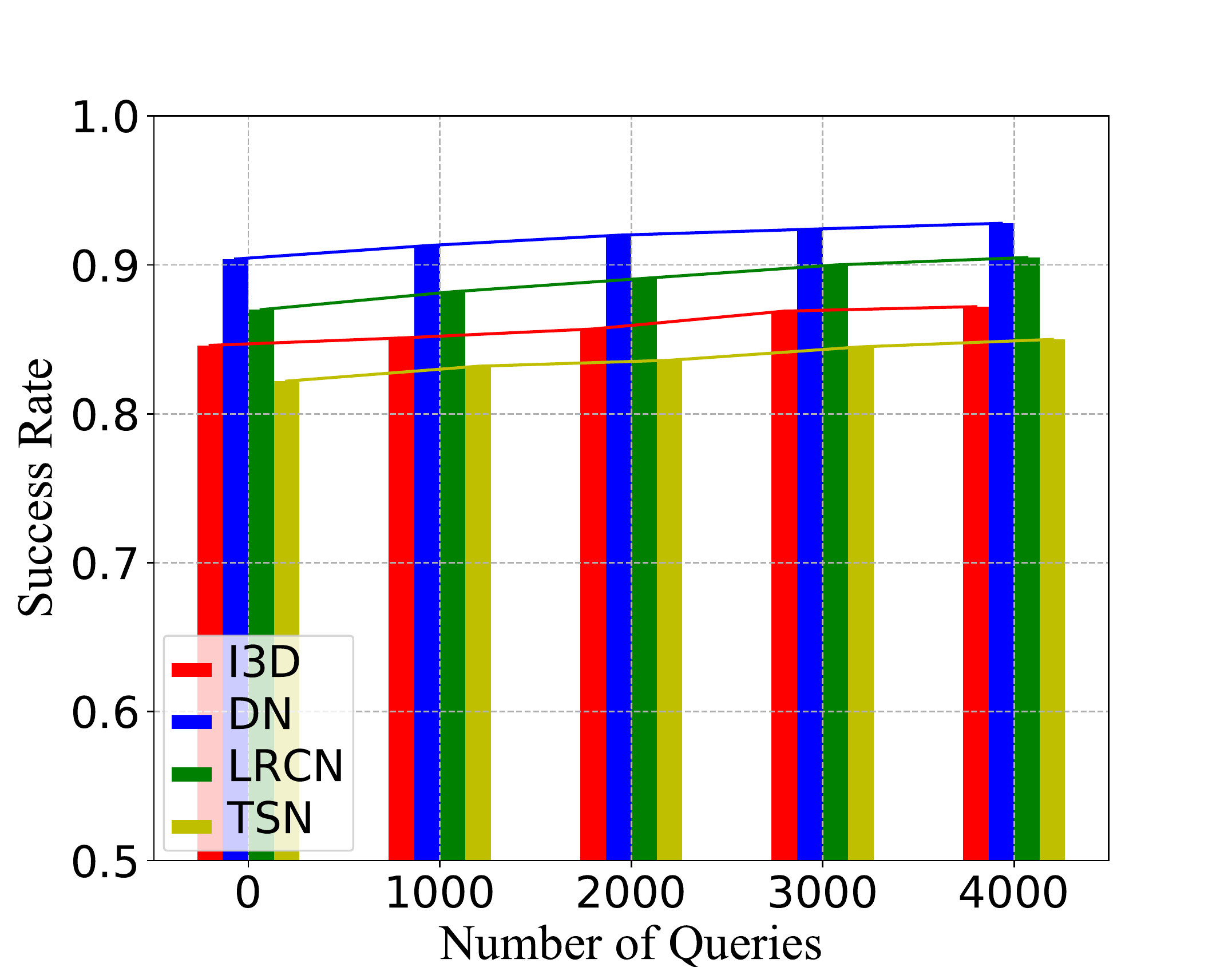} \label{fig:t-1} }\hspace{-0.2in}
		\subfigure[U3D$_g$]{
		\includegraphics[angle=0, width=0.495\linewidth]{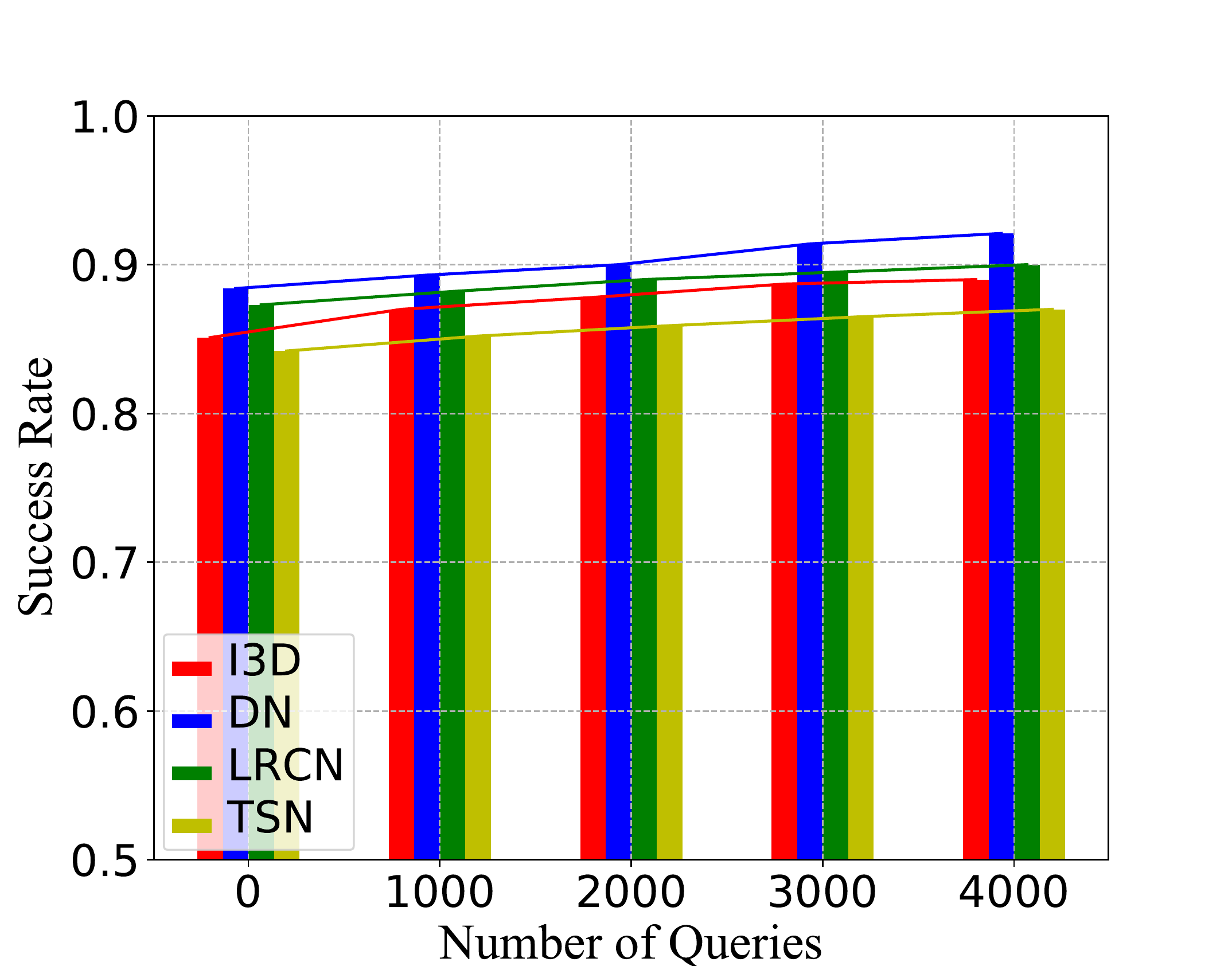} \label{fig:t-2} }\hspace{-0.2in}
    \vspace{-0.1in}
	\caption[Optional caption for list of figures]
	{Hybrid black-box attack performance (surrogate C3D).}\vspace{-0.1in}
	\label{fig:hybrid-c3d}
\end{figure}

To evaluate the hybrid attack, we choose the C3D as the surrogate model, and I3D, DN, LRCN, TSN as target models, respectively. Then, we follow the setting as the previous experiments on the UCF101 dataset (10\% for learning U3D while 90\% for target set) and U3D parameters. For the hybrid attack, we set the size of querying dataset as $50$ (randomly chosen), $\epsilon=8$ and $\omega=10$, vary the number of queries as \{0, 1000, 2000, 3000, 4000\} (``0'' means the original U3D attack without queries). Then, we apply PSO to optimize U3D perturbations on Equation \ref{eq:opt}, and report the success rate for both U3D perturbations against the four target models. 
Figure \ref{fig:hybrid-c3d} shows that the success rates of both U3D perturbations slightly increase as the number of queries increases. The hybrid attack with additional queries to the target model can improve the transfer-based attack to some extent \cite{usenix20-transfer}.

\subsection{Visual Impact and Human-Imperceptibility}
\label{sec:visual}

\noindent\textbf{Visual Impact}. We arbitrarily select two videos, ``shooting'' and ``fighting'' to demonstrate the visual differences of the adversarial examples. Figure \ref{fig:8} in Appendix \ref{app:add} presents a sequence of selected frames in two videos, and we can observe that the videos perturbed by U3D$_p$ and U3D$_g$ are much more human-imperceptible than C-DUP.

\vspace{0.05in}

\noindent\textbf{Human-Imperceptibility Study}.
We also conducted a human-imperceptibility study (with an IRB exempt) to validate if the perturbed videos could be visually discerned by humans compared with the original videos. We distributed the videos (original videos, adversarial examples based on U3D$_p$, U3D$_g$ and C-DUP) and an online survey to 87 anonymous students.  

Specifically, we randomly pick 870 videos from the UCF101 and UCF Crime datasets. To avoid bias on the distribution of adversarial examples, we randomly choose 435 videos to inject perturbations (218 videos by U3D$_p$ and 217 by U3D$_g$; all the adversarial examples are selected from the previous groups of experiments to successfully fool the video classifiers). The students do not know the distribution of adversarial examples. Then, 435 pairs of videos (the adversarial examples and their clean videos) are mixed with another 435 pairs of videos (the remaining 435 clean videos and their duplicates).

Next, we randomly assign 10 pairs of videos to each student. Each of the 87 students is expected to watch and annotate 10 pairs of videos (``visual difference'' or ``no visual difference''). We also ask the participants to identify the region of difference if they choose ``visual difference''. Finally, we collected 740 valid annotations on pairs of videos (74 students have submitted their results over a 1-week time frame; among the assigned videos to these students, there exist 386 adversarial examples). We found that 96.4\% (372) out of such 386 annotations are ``no visual difference'', while only 3.6\% (14 annotations) of them are annotated with ``visual difference''. Among the remaining 354 pairs of original videos and their duplicates, there also exist 7 annotations submitted as ``visual difference''. 

Meanwhile, we conduct the same group of studies for C-DUP by randomly picking another 870 different videos to inject perturbations (in order to avoid the bias on mutually linking the adversarial examples of U3D and C-DUP perturbed on the same videos).\footnote{If any visual difference is identified by the student from the C-DUP perturbed video (or vice-versa), this may give him/her prior knowledge to identify visual difference from U3D (or vice-versa) since both are perturbed.} As high as 85.4\% (304 out of 356 valid adversarial examples of C-DUP) can be successfully identified by the same group of students. All the above results have indicated that our U3D perturbations achieve high human-imperceptibility (significantly better than C-DUP).

\subsection{U3D against Defense Schemes}\label{exp:defend}

To our best knowledge, there are very few defense schemes against the adversarial attacks on videos (mostly on images). We comprehensively evaluate the performance of U3D against three major categories of state-of-the-art defense schemes, which are adapted towards video/U3D defenses. 
They include: (1) adversarial training \cite{madry2018towards, shafahi2020universal}; (2) adversarial example detection \cite{xiao2019advit,yin2019adversarial}; (3) certified robustness \cite{cohen2019certified,LecuyerAG0J19}. 

\vspace{0.05in}

\noindent\textbf{Attack and Defense Setting}. We use the U3D$_p$ and U3D$_g$ perturbations generated in Section \ref{exp:effect} (surrogate C3D and HMDB51 dataset) to craft adversarial examples on a dataset (e.g., UCF101 or UCF Crime). The adversarial examples will be used to attack the target model (C3D or I3D), which will be adversarially trained, integrated into the detection schemes, or certified with randomization. In all the tables in this subsection, ``Model'' refers to the target model, and ``Dataset'' refers to the dataset used to
craft adversarial examples. 

\vspace{0.05in}
\noindent\textbf{Adversarial Training}. Adversarial training \cite{GoodfellowSS14, madry2018towards, shafahi2020universal, shafahi2019adversarial} refers to the model training by adding adversarial examples into the training dataset. It has been empirically validated to be effective on improving the robustness against adversarial examples and maintaining accuracy on clean data. 

First, due to the universality of U3D, we evaluate U3D attack on a universal adversarial training (denoted as ``UAT'') \cite{shafahi2020universal} which defends against universal perturbations. Specifically, such scheme adopts PGD-based adversarial training \cite{madry2018towards} to formulate a min-max optimization problem as below:

\begin{equation}
\small
    \min_{\theta} \max_{\xi}: \frac{1}{|X|}\sum_{(x_i,y_i)\in X}L(\theta; x_i+\xi,y_i) ~~s.t.~~ ||\xi||_\infty\leq \epsilon
    \label{eq:madry}
\end{equation}

where $\theta$ denotes the model parameters, $X=\{(x_i, y_i), i\in[1, |X|]\}$ is the training sample set, $L(\cdot)$ is the loss function, and the $\ell_p$-norm of universal perturbation $\xi$ is bounded by $\epsilon$. Different from the conventional PGD-based adversarial training (computing the perturbation for each instance), the inner optimization problem seeks a universal (more precisely, batch $X$-universal) perturbation $\xi$ to maximize the adversarial loss w.r.t. the sample set $X$. It has been shown to be effective against universal perturbations compared to PGD-based adversarial training \cite{madry2018towards}. In addition, it is more efficient to compute one universal perturbation across all the training iterations, i.e., only updating perturbation $\xi$ once for each step \cite{shafahi2020universal}. 

Second, besides ``UAT'', we tailor the universal adversarial training towards U3D (denoted as ``U3D-AT''), and evaluate the U3D under a \emph{stronger defense setting} (see the white-box defense of G3 in Section \ref{sec:design}). Specifically, the defender knows the U3D function $\mathcal{N}(\cdot)$ but does not know the specific values of the U3D parameters $s$. Recall that the U3D perturbation is computed by optimizing Equation \ref{eq:u3d_opt_total}, which is formulated as a attack fitness function $\mathcal{A}(f, V, s)$. Then, we can also adapt the UAT framework by replacing the inner optimization objective with the U3D function as $\mathcal{A}(f, X, s)$: 

\begin{equation}
    \min_{\theta} \max_s:  \mathcal{A}(f,X,s) ~~ s.t. ~~ \xi=\mathcal{N}(T;s),~||\xi||_\infty\leq \epsilon \\    
    \label{eq:uni3d}
\end{equation}

where the norm-bounded U3D perturbation $\xi$ can be computed by the U3D function with the parameters $s$. Similar to UAT, we can iteratively update the best U3D perturbation among a batch of data ($X$) in the inner loop via $\mathtt{NoiseOpt}$, which adapts PSO to find the optimal parameters. 

For the experiments, we evaluate the defense performance of standard PGD-based adversarial training (denoted as ``Normal"), universal adversarial training (``UAT'') and our U3D-adaptive AT (``U3D-AT'') against our U3D perturbations, respectively. We split the datasets (UCF101 and UCF Crime) into the training dataset (80\%) and the test dataset (20\%). We set the perturbation bound $\epsilon=8$. For both UAT and U3D-AT, we set the batch size as 200. For UAT, we utilize FGSM to update $\xi$, and Momentum SGD optimizer to update model parameters as the original setting \cite{shafahi2020universal}. For the adversarially trained models, we evaluate the accuracy (Clean ACR) -- the predication accuracy on the clean data, besides the attack success rate (SR) for misclassification. Note that we also report the accuracy and SR of the normal models.

Table \ref{tab:at1} summarizes the results of the adversarial training against U3D on the UCF101. The accuracy of both UAT and U3D-AT on the clean data declines since the training has included adversarial examples. Nevertheless, the success rates of both U3D$_p$ and U3D$_g$ have been reduced against both UAT and U3D-AT. The U3D-AT performs better than the UAT, e.g., the attack SR of U3D$_p$ is 42.7\%$<$67.4\% on the C3D. This is because U3D-AT directly optimizes the defense on U3D (with the attack fitness function), which thus makes the model more robust against U3D. However, such U3D-AT is more like ``white-box'' defense in which the defender (model owner) already knows the adversary's strategy (e.g., U3D format and attack function). In practice, the defender usually cannot readily obtain such information.

\begin{table}[h]
 \centering\small
 \caption{Adversarial training on UCF101. See similar results on UCF Crime in Table \ref{tab:atcrime} (Appendix \ref{app:add}).}
 
  \begin{tabular}{c|c|c|cc}
  \hline
  
 \multirow{2}{*}{Model}&\multirow{2}{*}{Defense} &{Clean}&{U3D$_p$} &{U3D$_g$}   \\
 && ACR&(SR)&(SR)\\
  \hline
  
  \multirow{3}{*}{C3D} & Normal& 86.2\% & 83.7\% & 84.2\%
  \\&UAT  & 78.5\% & 67.4\% & 65.5\% \\ 
    &U3D-AT  &  77.2\% & 42.7\% & 45.3\% \\ 
      \hline

  \multirow{3}{*}{I3D}&  Normal& 88.7\% & 82.1\%& 82.6\%\\
 
  & UAT  &   80.4\% & 70.2\% &69.5\%\\
    &U3D-AT   &  78.6\% & 50.3\% &47.4\%\\
  \hline
  
  \end{tabular}\vspace{-0.1in}
  \label{tab:at1}
\end{table}

\vspace{0.05in}
\noindent\textbf{Adversarial Examples Detection}. Most detection schemes \cite{meng2017magnet,pmlr-v97-roth19a,liao2018defense,xiao2019advit,yin2019adversarial} locally train a detector or utilize feature characteristics in adversarial examples to determine if the input is perturbed or not. For instance, a detector can be trained on both clean data and adversarial examples via adversarial training \cite{yin2019adversarial}. Although detection schemes have difficulties on mitigating adversarial attacks (e.g., Magnet \cite{meng2017magnet} was broken by \cite{carlini2017magnet}, and some recent defenses were broken by adaptive attacks \cite{tramer2020adaptive}), we still evaluate our U3D against detection schemes (including that adapted to U3D). Note that the U3D attack can be both online and offline. Then, we evaluate both of them against the detection schemes (assuming that the offline detection can be executed with arbitrary runtime).

First, for the online detection, we choose AdvIT \cite{xiao2019advit} which is effective against the existing adversarial attacks on \emph{real-time} video recognition. It finds the inconsistency among the temporally close frames with the optimal flow information, assuming that perturbations can destroy the frame consistency to some extent. Specifically, given one target frame (to be detected), AdvIT first estimates the optimal flow between the target frame and previous $k$ frames, and then reconstructs pseudo frames by applying the optical flow to the beginning frame. Finally, it would compute the inconsistency score $c$ between the target frame and pseudo frames, where high inconsistency score indicates that the target frame is adversarial. To defend against the adaptive attacks, AdvIT applies the Gaussian noise to fuzz the optical flow for generating the pseudo frames. 

In the experiments, we randomly select 200 clean videos from the UCF101 and UCF Crime datasets (100 each), and apply both U3D$_p$ and U3D$_g$ perturbations to craft adversarial examples. We set the perturbation bound $\epsilon=8$. For detection, we set $k=3$ (which only slightly affects the detection rate) and utilize FlowNet \cite{Ilg_2017_CVPR} as the optical flow estimator in AdvIT. Then, we randomly select 5 frames in each video as the target frames, and average the inconsistency scores (reporting detection when $\geq 1$) to derive the detection results. 

Table \ref{tab:detect1} summarizes the detection accuracy (DR) and false positive rate (FPR) of AdvIT. It shows that U3D can bypass the detection of the state-of-the-art detection scheme, even though AdvIT achieves low false positive rates. For instance, AdvIT only obtains 12\% accuracy to detect U3D$_p$-based adversarial examples for the C3D. The results show that U3D is immune to the temporal consistency detection by AdvIT, since the U3D perturbations are constructed on continuous 3-dimensional noise, which can still retain the consistency in temporal space.

\begin{table}[h]
 \centering
 \caption{Detection and false positive rates of AdvIT \cite{xiao2019advit} 
 } 
  \small
  \begin{tabular}{c|c|cc|cc}
  \hline
  
 \multirow{2}{*}{Model}& \multirow{2}{*}{Dataset} &\multicolumn{2}{c|}{U3D$_p$}& \multicolumn{2}{c}{U3D$_g$}  \\ \cline{3-4}\cline{5-6}
   & & DR &FPR &DR &FPR \\
  \hline
  
  \multirow{2}{*}{C3D} & UCF101 &12\%  & 2\%&18\% & 2\% \\
   & UCF Crime & 12\%  & 5\%&19\% & 3\% \\
      \hline

  \multirow{2}{*}{I3D} & UCF101& 10\% & 3\% &17\%&3\% \\
  & UCF Crime & 12\%  & 5\%&22\% & 3\% \\
  \hline
  \end{tabular}\vspace{-0.15in}
  \label{tab:detect1}
\end{table}

\begin{table}[h]
 \centering
 \caption{Detection AUC of AdvIT \cite{xiao2019advit} against U3D, C-DUP, V-BAD, and H-Opt. C3D:1st/3rd row. I3D:2nd/4th row} 
  \small
  \begin{tabular}{c|c|c|c|c|c}
  \hline
  
 {Dataset}& {U3D$_p$}& {U3D$_g$}& C-DUP & {V-BAD} & {H-Opt} \\ 
  \hline
  
  \multirow{2}{*}{UCF101}&  54.2\%  & 56.7\%& 97.2\%&98.4\% & 99.2\% \\ 
    & 56.4\% & 55.3\% &98.7\%& 97.3\% & 98.6\%\\
      \hline

  \multirow{2}{*}{UCF Crime}&  61.2\%  & 64.8\%& 97.6\%&99.5\% & 98.3\% \\ 
    & 55.6\% & 58.1\% &97.4\% &99.7\% & 99.8\%\\
  \hline
  \end{tabular}\vspace{-0.05in}
  \label{tab:detect-advit}
\end{table}

Furthermore, we have evaluated the Area Under Curve (AUC) values of the Receiver Operation Characteristic Curve (ROC) of AdvIT for U3D perturbations and other three benchmarks: C-DUP \cite{LiNPSKRS19}, V-BAD \cite{jiang2019black} and H-Opt \cite{abs-1911-09449}. The AUC metric represents the probability that the detector assigns a higher score to a random positive sample (adversarial example) than to a random negative sample (clean data) \cite{xiao2019advit}. It can better measure the detection performance than the DR/FPR. Table \ref{tab:detect-advit} summarizes the results. From the table, we can observe that the AUC values of U3D are close to random guess, e.g., 54.2\% (U3D$_p$) and 56.7\% (U3D$_g$) on C3D and UCF101 while all the benchmarks can be almost fully detected by AdvIT (\emph{all the AUC values are very close to 1}). This occurs since the temporal consistency cannot hold in the adversarial examples by C-DUP, V-BAD and H-Opt (perturbations are generated specific to the frames as frame-by-frame perturbations).

Second, for the offline detection, we evaluate the U3D against another recent work \cite{yin2019adversarial} based on the adversarial training \cite{madry2018towards}. If the universal adversarial training (UAT) can defend against U3D to some extent, we can also extend it to train a universal perturbation detector against U3D. Specifically, the \emph{asymmetrical adversarial training} (AAT) \cite{yin2019adversarial} trains $K$ detectors (for a $K$-class classification model) to detect adversarial examples. Given an input $x$, each detector $h_k, k\in[1,K]$ will output a \emph{logit} score corresponding to the class label, which can determine if data is perturbed or not (see details in \cite{yin2019adversarial}). To defend against the U3D, we revise the $K$ detectors $h_k, k\in[1,K]$ with the UAT by changing the training objective as below (denoted as ``U3D-AAT''):

\vspace{-0.1in}

\begin{align}\small
    \min_{\theta}: & ~~[\mathop{\mathbb{E}}_{x\sim D_{k}'}\max_s L(h_k(x+\xi), 1)+\mathop{\mathbb{E}}_{x\sim D_k}L(h_k(x),0)] \nonumber\\
    s.t. &~~ \xi=\mathcal{N}(T;s),~||\xi||_\infty\leq \epsilon 
\end{align}

The objective includes two parts: (1) the maximum loss of adversarial examples (by U3D perturbation $\xi$) on the out-of-class data samples $D_{k}'$; (2) the loss of intra-class natural data samples $D_k$. $L(\cdot)$ is a loss function, e.g., binary cross-entropy loss. For the inner optimization of the first part, we adopt similar procedures as U3D-AT to update the U3D perturbations (as depicted earlier).

\begin{table}[h]
 \centering
 \caption{Detection and false positive rates of U3D-AAT.} 
  \small
  \begin{tabular}{c|c|cc|cc}
  \hline
  
 \multirow{2}{*}{Model}& \multirow{2}{*}{Dataset} & \multicolumn{2}{c|}{U3D$_p$}& \multicolumn{2}{c}{U3D$_g$}   \\ \cline{3-4}\cline{5-6}
   & & DR &FPR    &  DR &FPR  \\
  \hline
 
  \multirow{2}{*}{C3D}& 
     UCF101 &56.2\%  & 6.3\%&53.4\% & 5.9\% \\
     &HMDB51 &44.5\%  & 8.2\%&47.4\% & 7.1\% \\
      \hline

  \multirow{2}{*}{I3D}& UCF101 & 55.4\% &4.2\% &56.5\%&5.1\% \\
  &HMDB51 &52.6\%  & 5.7\%&54.3\% & 5.9\% \\
  \hline
  \end{tabular}\vspace{-0.1in}
  \label{tab:aat}
\end{table}

To evaluate the performance of the detectors, we choose the action classification on the UCF101 and HMDB51 as $K$-Class problem ($K=101$ and $51$). Specifically, we split the training/testing datasets by 80\%/20\% for each category. We set the perturbation bound $\epsilon=8$, and apply the two U3D perturbations to craft the adversarial examples, which are mixed up with the clean videos for detection (for instance, in UCF101 dataset, there are 2664 clean videos, 2664 videos perturbed by U3D$_p$, and 2664 videos perturbed by U3D$_g$). We adopt the \emph{integrated classifier} which computes the estimated class label $c=f(x)$ with the original classifier $f$ and computes a logit vector $h_c(x)$ using the corresponding detector $h_c$ \cite{yin2019adversarial}. We report the detection accuracy (DR) and false positive rate (FPR). The results in Table \ref{tab:aat} have shown that such universal adversarial detector can detect the U3D perturbations to some extent: the universal AAT detector can achieve 
about 50\% detection rate while maintaining a low FPR (less than 7\%). Such FPR is reasonable considering there could still exist overlapped adversarial subspaces, i.e., U3D-AAT may not be trained to be perfect to learn U3D perturbations and thus separate the perturbed video and clean ones. However, training such AAT detectors should know the U3D attack (\emph{white-box defense}), and it is only limited to defend against offline attacks due to the computational costs.

\vspace{0.05in}

\noindent\textbf{Certified Robustness}. Recently, certified schemes \cite{LecuyerAG0J19, cohen2019certified,kumar2020curse,JMLR:v21:20-209,wong2018provable} have been proposed to defend against norm-bounded adversarial perturbations with theoretical guarantees. We evaluate the U3D attack against two representative certified schemes: PixelDP \cite{LecuyerAG0J19} and randomized smoothing \cite{cohen2019certified}.

First, PixelDP \cite{LecuyerAG0J19} adopts the Gaussian mechanism of differential privacy to slightly randomize the image pixels \cite{wang20pets}. After injecting Gaussian noise, the small change of image pixels (adversarial perturbation) will not affect the classification results with some probabilistic bound (thus provide robustness guarantee for DNN models). It will be extended from protecting image DNN models to video DNN models.

To evaluate the U3D attack against PixelDP, we modify the video DNN models by placing the noise layer in the first convolutional layer under the same Gaussian mechanism setting \cite{LecuyerAG0J19} w.r.t. an $\ell_2$ attack bound $L=0.1$ (such setting ensures a high accuracy in \cite{LecuyerAG0J19}). We split training/test as 80\%/20\% for retraining the model.  Note that PixelDP admits that the certified effectiveness against $\ell_\infty$ attacks is substantially weaker via empirical evaluations (which conforms to the performance of other certified schemes such as randomized smoothing). Then, we generate U3D perturbations bounded by $\ell_2$ norm value of $0.5$ (which indeed generates very minor perturbations in case of very high video dimensions). 

\vspace{-0.1in}

\begin{table}[h]
 \centering
 \caption{Accuracy (ACR) and success rate (SR) of PixelDP \cite{LecuyerAG0J19} (UCF101 and UCF Crime). 
 }
 \small
 
  \begin{tabular}{c|c|c|cc}
  \hline
 
  \multirow{2}{*}{Model}&\multirow{2}{*}{Dataset} &{Clean}&{U3D$_p$} &{U3D$_g$}   \\
 && ACR&(SR)&(SR)\\
  \hline

  \hline
  
  \multirow{2}{*}{C3D} & UCF101& 63.2\% & 83.4\% & 85.3\% \\ 
  & UCF Crime& 65.9\% & 86.2\% & 89.7\%\\
      \hline

  \multirow{2}{*}{I3D}&UCF101&65.8\% & 82.3\% &79.4\%\\
   &UCF Crime& 67.4\% & 84.7\% &85.2\%\\
  \hline
  
  \end{tabular}\vspace{-0.05in}
  \label{tab:defend1}
\end{table}

We report the classification accuracy of PixelDP on clean videos, and the success rates of the U3D attack in Table \ref{tab:defend1}. The accuracy of PixelDP drastically declines after injecting Gaussian noises (vs. the baseline models), e.g., 86.2\%$\rightarrow$63.2\% on C3D. Meanwhile, the U3D attack can still achieve high success rates in all the cases. This shows that PixelDP cannot defend against U3D since PixelDP only ensures a weak bound with the Gaussian mechanism of differential privacy. 

Second, we also evaluate the certified robustness via randomized smoothing \cite{cohen2019certified}. It provides a tight guarantee (based on the Neyman-Pearson Lemma) for any random classifier by smoothing inputs with an additive isotropic Gaussian noise. However, it only certifies $\ell_2$ radius of the perturbation bound. The certified schemes via smoothing against $\ell_\infty$ have been shown to be ineffective as the input dimensionality $d$ increases. The certified radius is bounded by $O(\frac{1}{\sqrt{d}})$ as $p>2$ \cite{kumar2020curse,JMLR:v21:20-209}.

To evaluate our U3D attack against such certified scheme, we first generate the optimal U3D perturbations $\xi$ by changing perturbation bound $\ell_\infty$ in $\mathtt{NoiseOpt}$ to $\ell_2$.\footnote{Since randomized smoothing cannot certify defense against $\ell_\infty$ bounded attack for high dimensional inputs (e.g., videos) \cite{kumar2020curse,JMLR:v21:20-209}, the U3D perturbations using $\ell_2$ bound instead of $\ell_\infty$ are still effective against such scheme.} Specifically, we evaluate the accuracy of the smoothing classifier on the perturbed videos against U3D, which is the percentage of the perturbed videos to be correctly classified (we also evaluate the accuracy on the clean videos as benchmarks). Furthermore, we also derive certificated radius $R$ for the videos, which indicates that the classification results can be certified against any perturbation with $\ell_2$-norm no greater than $R$ (see \cite{LecuyerAG0J19}). 

Next, we set the number of Monte Carlo samples as $n=100/1000$ and failure rate $\alpha=0.001$. The failure rate indicates that the robust classifier can have 1-$\alpha$ confidence to return the classification result. We set the Gaussian variance $\sigma=0.25$ (same as \cite{LecuyerAG0J19}), and the radius bound for U3D perturbations as $\epsilon=0.5$ (which generates minor perturbations in case of high video dimensions). We report the accuracy and average certified radius in Table \ref{tab:certified1}. The results show that the randomized smoothing cannot defend against the U3D attack under $\ell_2$-norm perturbations (can only certify very small radius), and the robust classifier only achieves less than $70\%$ accuracy on the clean video samples. 

\vspace{-0.05in}

\begin{table}[h]
 \centering
 \caption{Accuracy and radius of rand. smoothing  \cite{cohen2019certified} on UCF101. See similar results in Table \ref{tab:certified2} (Appendix \ref{app:add}).} 
  \small
  \begin{tabular}{c|c|c|cc|cc}
  \hline
  
 \multirow{2}{*}{Model} &\multirow{2}{*}{n} &{Clean} & \multicolumn{2}{c|}{U3D$_p$} & \multicolumn{2}{c}{U3D$_g$}   \\ \cline{4-5}\cline{6-7}
   & &ACR &ACR & Radius    &  ACR & Radius  \\
  \hline
  
  \multirow{2}{*}{C3D}& 100&67.4\% &  14.5\% & 0.23 &15.2\% & 0.19 \\ 
    & 1000 & 68.2\%  &  16.2\% & 0.24 &18.4\% & 0.25 \\ 
      \hline

  \multirow{2}{*}{I3D}& 100& 71.5\% &  21.7\% & 0.32 &19.8\% & 0.28 \\ 
    & 1000 & 72.2\%  &  25.2\% & 0.37 &21.2\% & 0.26 \\
  \hline
  
  \end{tabular}\vspace{-0.15in}
  \label{tab:certified1}
\end{table}

\subsection{Practicality for the U3D Attack}
\label{sec:real}

We now discuss the possible attack vectors, and evaluate the U3D attack on a real system. Prior real-time video attack scenarios can also be used for U3D (e.g., manipulating the system via a pre-installed malware in C-DUP \cite{LiNPSKRS19}). Besides them, we design extra ways for the real-time attack. Other adversarial attacks on videos (including future attacks) can also use our physical scenarios to inject real-time perturbations.

First, the network topology of the recent intelligent video surveillance system (VSS) \cite{costin2016security,obermaier2016analyzing}, include: (1) camera; (2) communication network; (3) server. Then, the adversary needs to inject the U3D perturbations in two cases: data-at-rest and data-in-motion \cite{MirskyMSE19}. The data-at-rest locally stores videos in the camera or the server. The data-in-motion transfers videos across the network or loads them into the volatile memory. Per the potential threats to VSS \cite{obermaier2016analyzing,IPCamera}, we consider the local infiltration to the systems in two scenarios: (1) malware; (2) man-in-the-middle (MITM) attack. First, malware can be locally installed via a malicious firmware update over the USB port. Moreover, the surveillance cameras could be sold through legitimate sales channels with the pre-installed malware \cite{malware16}. Second, for the MITM attack, the adversary could access to the local network (e.g., by penetration) which connects to the camera and server, and behave like a normal user. Here, we take the MITM attack as an example.

Specifically, we setup a local camera-server network, where one PC works as the surveillance camera to continuously send video streams to another PC (as a server) using the real-time streaming protocol (RTSP). Then, we use the third PC as the adversary running $\mathtt{Ettercap}$ (https://www.ettercap-project.org/) with ARP poisoning to implement the man-in-the-middle attack (sniffing the network traffic). All three PCs use Ubuntu 18.04 OS, connected on a LAN. According to the recent survey on the security of IP-based video surveillance systems \cite{shodan20, IPCamera}, a large number of unencrypted cameras (4.6 millions) are exposed to the network, e.g., using HTTP instead of HTTPS. Although the percentage of such unencrypted cameras is not disclosed, the unencrypted RTSP has been a major security vulnerability in video surveillance \cite{shodan20, IPCamera}. Thus, in our attack setting, we assume that the camera network is open with unencrypted RTSP. By exploiting the vulnerabilities, the adversary will target the camera-server communication without decryption and temporarily intercept the communication session by injecting the TEARDOWN request to the server. When the server tries to send a new request for the new communication session with the camera, the adversary will capture it, modify the delegated client port and forward the request to the camera. Finally, the adversary can receive video streams from the camera in real time. 

Note that we utilize the $\mathtt{FFmpeg}$ compiled with the video encoder $\mathtt{libx264}$ to execute the codec (decode and encode process) on the video from RTSP streams. Since our attack is performed through unencrypted video streams, there is no extra cost for decrypting video packets. To evaluate the computational overheads for the codec on the video streams, we set the following encoding parameters: (1) PRESET (encoding speed): ``medium'' by default; (2) bit rate: same as the streaming video bit rate ($\sim$ 350kbps) \cite{ffmpeg}. The average cost for the codec on the video is $\sim 0.3<1$ second, which will not affect the streaming video quality. It can also be accelerated by hardware, e.g., GPU. Overall, we have experimentally shown that the delay of our U3D attack (including both codec and injection) are negligible. Finally, the adversary can forward the perturbed video streams to the server for misclassifications. 

\begin{table}[htbp]
 \centering\small
 \caption{Amortized runtime (each frame) for attacking the streaming video on UCF101 (in Seconds). See similar results on UCF Crime in Table \ref{tab:time2} (Appendix \ref{app:add}).}  
  \begin{tabular}{c|cc|c}
  \hline
  
   Video Name   &  Codec & Inject  & Runtime  \\
  \hline
  
  Bowling   & 0.010 &0.004   & 0.014\\
  BoxingPunchingBag   & 0.012 &0.005        & 0.017 \\
  CliffDiving   &0.010 & 0.005  & 0.015      \\
  CuttingInKitchen & 0.009 &  0.005        &  0.014\\
  HorseRace   & 0.010 &0.004  & 0.014       \\
  \hline
  
  \end{tabular}\vspace{-0.05in}
  \label{tab:time}
\end{table}

\noindent\textbf{Evaluation}. We randomly pick 10 videos (5 videos from each of UCF101 and UCF Crime) to evaluate the attack against the video classification and anomaly detection. For each video, we repeat 10 times while injecting 10 different U3D perturbations (pre-generated with the HMDB51 dataset and C3D) in the video streams. The attack success rate for classification is 88\% (44/50) and for anomaly detection is 98\% (49/50). Table \ref{tab:time} presents the amortized online time for processing each frame. All the runtimes are less than 1/30 (the frame rate is between 24fps and 30fps in experimental videos). Thus, U3D can efficiently attack streaming videos with negligible latency.

\vspace{-0.2in}

\begin{figure}[!tbh]
	\centering
		\subfigure[Road Accident (U3D$_p$)]{
		\includegraphics[angle=0, width=0.49\linewidth]{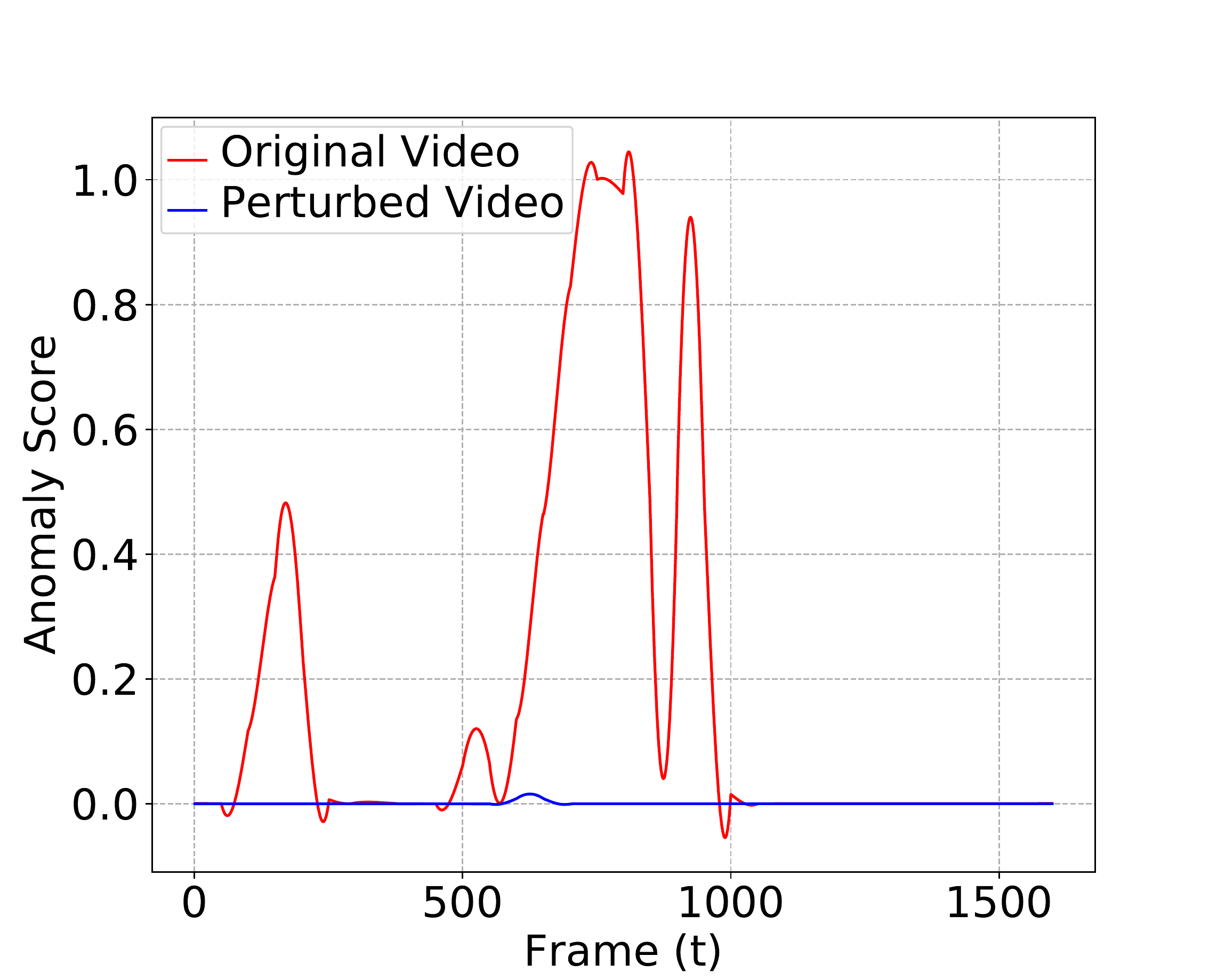} \label{fig:an-1} }\hspace{-0.2in}
		\subfigure[Explosion (U3D$_g$)]{
		\includegraphics[angle=0, width=0.49\linewidth]{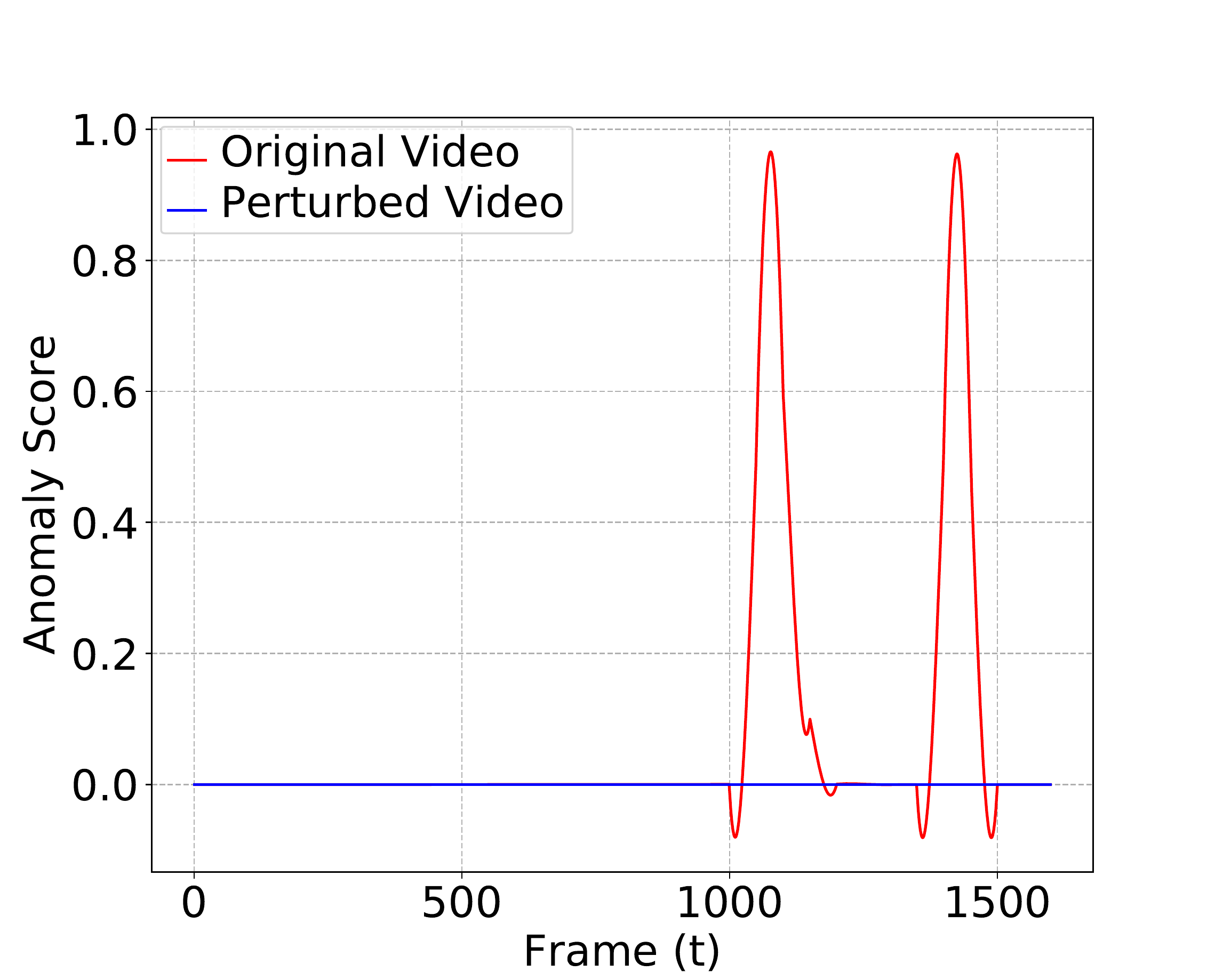} \label{fig:an-2} }\vspace{-0.1in}
	\caption[Optional caption for list of figures]
	{Real-time attack on anomaly detection}\vspace{-0.1in}
	\label{fig:anomaly}
\end{figure}

Moreover, Figure \ref{fig:anomaly} presents the real-time anomaly scores of two example videos (i.e., ``Road Accident'' and ``Explosion''), where each streaming video (sent from the camera to the server) is perturbed by a U3D perturbation in real-time (w.l.o.g., U3D$_p$ for ``Road Accident'' and U3D$_p$ for ``Explosion''). In Figure \ref{fig:an-1} (``Road Accident''), we can observe that there are three wave peaks in the original video, e.g., around frame 750, which will trigger the anomaly alarm (reporting ``Road Accidents'' if the score is greater than a pre-set threshold). While our U3D attack perturbs the streaming video, the anomaly scores of the perturbed video are reduced to almost zero in all the frames. The ``Explosion'' example (Figure \ref{fig:an-2}) also shows similar results. This illustrates that our U3D can perfectly compromise the video anomaly detection systems. 

\begin{table}[h]
 \centering\small
 \caption{Success rates of U3D perturbations (boundary effect-free), injected at 10 different times for each video.}
  \begin{tabular}{c|cc|cc}
  \hline
  
\multirow{2}{*}{\diagbox[width=6em]{Noise}{Model}} & \multicolumn{2}{c|}{C3D}&\multicolumn{2}{c}{I3D}\\\cline{2-3}\cline{4-5}& UCF101 &UCF Crime & UCF101 &UCF Crime\\
\hline
 {U3D$_p$}  &81.2\%&90.3\%&80.2\%&85.3\%\\
 {U3D$_g$} &84.5\% &93.0\% &82.6\%&89.4\%  \\\hline
 
  \end{tabular}
  \label{tab:taa}
\end{table}

Finally, to validate the boundary effect-free property of the U3D attack on streaming videos, we conduct another group of experiments on the UCF101 and UCF Crime datasets. Note that the lengths of videos are at least 15 seconds. Then, for each input video in two datasets, we insert the U3D perturbation at 10 different times (from 0 to 5s with a step of 0.5s), and the classification and anomaly detection will start from the first perturbed frame to the end of the video. Table \ref{tab:taa} summarizes the results for success rates in two applications. We can observe that our U3D perturbations can still achieve high success rates while the misalignment may occur, e.g., U3D$_p$ still achieves 81.2\% on UCF101 against C3D, and U3D$_g$ achieves 93.0\% on UCF Crime against C3D. This shows that U3D can mitigate the boundary effect well.

\section{Mitigation of U3D Perturbations}
\label{sec:disc}

The experiments show that the adversarial training (AT) is still the state-of-the-art on improving the model robustness regardless of overheads. The universal AT and AT adapted to U3D (U3D-AT) have shown some effectiveness on reducing the attack success rates (though the accuracy on clean videos has been reduced). To further improve the performance of AT against U3D, we can enhance the search in a larger U3D perturbation space. Also, we can integrate the adaptive inference method, e.g., applying stochastic interpolation to reduce the effect of U3D \cite{pang2019mixup}, and certified robustness \cite{cohen2019certified}.
 
For detection methods, the properties of the procedural noise (e.g., low frequency texture structure) can be utilized. For instance, since the background scenes in most surveillance videos captured by static cameras do not change, the defender can extract the static frame of the background and compare it with the perturbed video frame(s) to check the possible perturbation. An alternative way is to check the moving objects or humans in the videos. Since the U3D perturbations applied to the same object in different frames are likely to be different due to the changed coordinates, the deviations between the perturbed object in different frames might be identified. This needs other object detection/tracking algorithms, which may only be suitable for offline analysis due to high overheads. 

Furthermore, although certified robustness cannot defend against the U3D, it is promising since it provides theoretical guarantee against norm-bound perturbations. One potential method to improve the robustness is the integration of randomized smoothing with UAT, which could potentially make the trained model robust against more unknown perturbations and thus improve the robust accuracy. However, this also poses challenges on expensive training (not model-agnostic either). We should also address the high dimensionality of videos since the certified guarantee can be jeopardized drastically on high dimensional data under $\ell_\infty$ bound. Thus, we can execute transformation to reduce the dimension of input data (e.g., by autoencoder) while certifying the robustness after transformation. We will explore these in the future. 

Last but not least, we can also mitigate the online U3D attacks by enhancing the security of video recognition systems, e.g., upgrade to encrypted communication channels or add watermarking to the video streams (to detect injections).

\section{Related Work}
\label{sec:related}
Security in machine learning, especially the vulnerabilities of AI systems to the adversarial inputs, has been intensively studied in both security and machine learning communities. Since adversarial examples were introduced \cite{biggio2013evasion,SzegedyZSBEGF13, GoodfellowSS14}, there have been numerous works on attacking image classifiers. For instance, FGSM \cite{GoodfellowSS14}, PGD \cite{madry2018towards}, UAP \cite{Moosavi-Dezfooli17} and many others \cite{kurakin2016adversarial, Moosavi-Dezfooli16, carlini2017towards}, work well in the white-box setting. For black-box attacks, researchers have proposed two main types of methods: transfer-based \cite{liu2016delving,papernot2017practical,cheng2019improving} and query-based attacks \cite{chen2017zoo,brendel2017decision, pmlr-v80-ilyas18a,DBLP:conf/iclr/ChengLCZYH19}. Recently, a hybrid attack \cite{usenix20-transfer} combines both of them to improve the attack performance. Moreover, adversarial attacks emerge in voice recognition \cite{carlini2016hidden,chen2019real}, malware classification \cite{grosse2016adversarial}, text understanding \cite{LiJDLW19}, etc.

Recent research has extended adversarial attacks from attacking DNNs on 2-D images to 3-D videos \cite{wei2019sparse, LiNPSKRS19,abs-1911-09449,jiang2019black}. Wei et al. \cite{abs-1911-09449} proposed a heuristic algorithm based on the query-based optimization attack \cite{DBLP:conf/iclr/ChengLCZYH19} to search the saliency region in the video frames for perturbation. V-BAD \cite{jiang2019black} utilizes natural evaluation strategy (NES) \cite{pmlr-v80-ilyas18a} to query the target model for estimating gradient, and then craft adversarial examples via PGD. Both methods compute the perturbation for each frame, which requires heavy computational overheads. They cannot attack real-time videos (due to lack of universality either). C-DUP \cite{LiNPSKRS19} applies GAN to generate universal perturbations offline and attack real-time video classification. However, it is a white-box attack, which is also limited to only the C3D model. More importantly, due to lack of consistency in the perturbations across frames, all these three attacks can be directly mitigated by AdvIT \cite{xiao2019advit} with high accuracy.

To defend the model against adversarial attacks, a wide range of defense schemes \cite{madry2018towards,yin2019adversarial,cohen2019certified, LeCunMBCF05,shafahi2020universal, xiao2019advit} have been proposed, which aim to either improve the robustness of model or detect adversarial examples. 
To our best knowledge, existing defense schemes (e.g., \cite{madry2018towards,LecuyerAG0J19}) mainly work on images, and have not empirically studied videos. Instead, we have thoroughly evaluated our U3D attack by redesigning current defense schemes in Section \ref{exp:defend}, which show some effectiveness against the U3D attack on videos. We anticipate that our U3D can motivate to build more robust defense schemes for DNN-based video recognition.

\section{Conclusion}
\label{sec:concl}
In this paper, we have successfully constructed two novel U3D perturbations to universally attack multiple DNN-based video recognition systems in the black-box setting. The proposed U3D$_p$ and U3D$_g$ can be efficiently generated on-the-fly while ensuring transferability, universality and human-imperceptibility. Also, U3D$_p$ and U3D$_g$ can be applied to attack video streams in real-time applications. Furthermore, we have conducted extensive experiments on three large-scale video datasets to validate the performance of U3D perturbations by benchmarking with the state-of-the-art attacks. The experimental results demonstrate that our U3D attack greatly outperforms other attacks (e.g., C-DUP) on attack success rate, transferability, and human-imperceptibility. We also perform experiments to evaluate U3D attack against three different types of defense schemes, adapted to universal perturbation or U3D on videos. It is more difficult to defend against U3D compared to C-DUP, V-BAD, and H-Opt (e.g., by AdvIT).

\section*{Acknowledgements}

This work is partially supported by the National Science Foundation (NSF) under the Grants No. CNS-1745894 and CNS-2046335. We would like to thank Zhaorui Liu and Junwen Chen for their help on some preliminary results and figures. We are also grateful to the anonymous reviewers and the PC point of contact for their constructive comments.

%\bibliographystyle{IEEEtranS}

% Generated by IEEEtranS.bst, version: 1.14 (2015/08/26)

\appendix

\label{sec:apdx}

\subsection{Additional Experimental Results}\label{app:add}

\begin{table}[htbp]
 \centering\small
 \caption{Transferability: transfer rate (TR) on HMDB51 from surrogate model $f_{srg}$ to target model $f_{tar}$}
 \begin{tabular}{c|c|ccccc}
  \hline
Noise &\diagbox[width=4em]{$f_{srg}$}{$f_{tar}$} &C3D&I3D& DN & LRCN  & TSN\\  
  \hline
   \multirow{5}{*}{U3D$_p$}&C3D&--&92.0\%  & 89.5\% &83.2\% &84.6\%\\
   &I3D&87.6\%  &--& 91.2\% &82.5\% &81.4\%\\
   &DN&82.3\%  &81.6\% & --& 84.5\% &80.5\%\\
   &LRCN&85.0\%  &85.4\%&95.3\%&-- &85.6\%\\
   &TSN &82.5\% & 85.7\%  & 88.0\% &84.2\% &--\\
   \hline
   \multirow{5}{*}{U3D$_g$}&C3D&--&86.6\% &  96.4\%  & 81.4\% & 83.1\%\\
  &I3D&92.2\%  &  --&  96.0\%  & 88.3\% & 84.5\%\\
  &DN&82.0\%  &84.8\% & --& 87.5\% &82.3\%\\
   &LRCN&85.6\%  &83.4\%&88.2\%&-- &82.9\%\\
   &TSN &84.6\% & 81.2\%  & 86.1\% &84.2\% &--\\
  
  \hline
  
  \end{tabular}
  \label{tab:cross-model1}
\end{table}

\begin{table}[htbp]
 \centering\small
 \caption{Transferability: transfer rate (TR) on UCF Crime from surrogate model $f_{srg}$ to target model $f_{tar}$}
 \begin{tabular}{c|c|ccccc}
  \hline
Noise &\diagbox[width=4em]{$f_{srg}$}{$f_{tar}$} &C3D&I3D& DN & LRCN  & TSN\\  
  \hline
   \multirow{5}{*}{U3D$_p$}&C3D&--&91.5\%  & 94.1\% &92.3\% &89.0\%\\
   &I3D&90.7\%  &--& 94.5\% &90.2\% &89.1\%\\
   &DN&87.2\%  &87.4\% & --& 88.2\% &90.7\%\\
   &LRCN&92.8\%  &87.2\%&92.4\%&-- &86.1\%\\
   &TSN &91.7\% & 90.2\%  & 93.4\% &91.7\% &--\\
   \hline
   \multirow{5}{*}{U3D$_g$}&C3D&--&87.0\% &  93.2\%  & 86.3\% & 85.3\%\\
  &I3D&91.3\%  &  --&  93.4\%  & 90.2\% & 89.0\%\\
  &DN&89.3\%  &88.4\% & --& 93.4\% &89.5\%\\
   &LRCN&93.2\%  &90.8\%&91.2\%&-- &90.4\%\\
   &TSN &89.4\% & 88.6\%  & 92.4\% &89.5\% &--\\
  \hline
  
  \end{tabular}
  \label{tab:cross-model2}
\end{table}

\begin{table}[htbp]
 \centering\small
 \caption{Universality (success rate (SR); I3D surrogate).}
  \begin{tabular}{c|c|ccccc}
  \hline
 
Noise &\diagbox[width=6em]{$X$}{$Y$} & HMDB51 & UCF101 &  UCF Crime \\
  \hline
   &HMDB51 & {89.6}\% & 80.4\% & 87.1\%\\
    U3D$_p$ &UCF101 & 83.5\% & {86.3}\% & 93.4\%\\
     &UCF Crime & 80.1\% & 82.4\% & {96.0}\%\\\hline
   &HMDB51& {86.3}\% & 82.9\% & 90.5\%\\
     U3D$_g$ &UCF101 & 82.5\% & {86.7}\% & 91.7\%\\
     &UCF Crime & 80.1\% & 84.3\% & {98.5}\%\\
    
  \hline
  
  \end{tabular}
  \label{tab:cross-dataset1}
\end{table}

First, Table \ref{tab:cross-model1} and \ref{tab:cross-model2} show additional results for cross-model transferability tested on the HMDB51 and UCF Crime datasets. Table \ref{tab:cross-dataset1} show the results for cross-data universality by considering I3D as the surrogate DNN to learn the U3D. All the results have confirmed the high transferability and universality of U3D (similar to the results in Section \ref{exp:transfer}).

\begin{figure*}[!tbh]
	\centering

	\subfigure[U3D$_p$]{
	\hspace{-0.1in}
   	 	\includegraphics[width=0.12\textwidth]{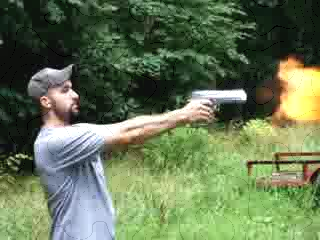}
   	 	\includegraphics[width=0.12\textwidth]{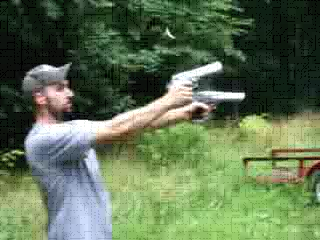}
   	 	\includegraphics[width=0.12\textwidth]{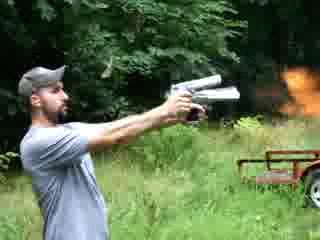}
   	 	\includegraphics[width=0.12\textwidth]{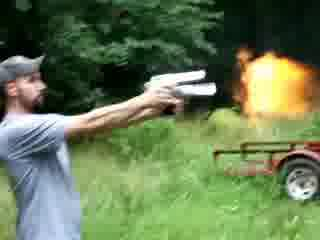}
     	\includegraphics[width=0.12\textwidth]{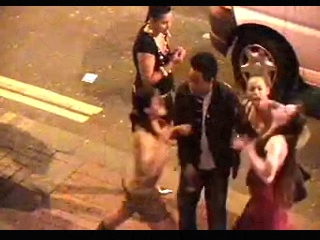}
   	 	\includegraphics[width=0.12\textwidth]{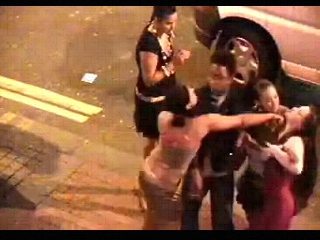}
   	 	\includegraphics[width=0.12\textwidth]{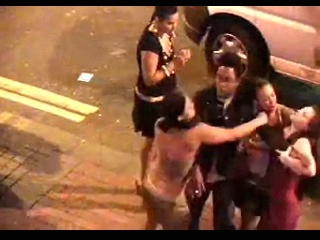}
   	 	\includegraphics[width=0.12\textwidth]{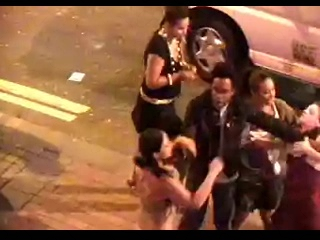}
	\label{fig:8_b}
	}
	\subfigure[U3D$_g$]{
	\hspace{-0.1in}
   	 	\includegraphics[width=0.12\textwidth]{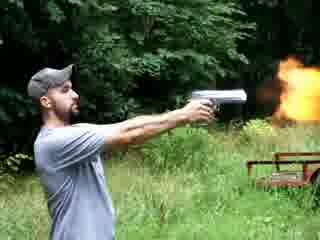}
   	 	\includegraphics[width=0.12\textwidth]{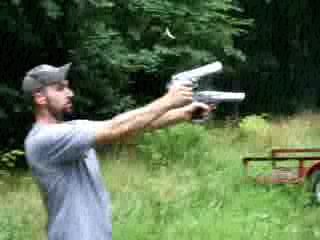}
   	 	\includegraphics[width=0.12\textwidth]{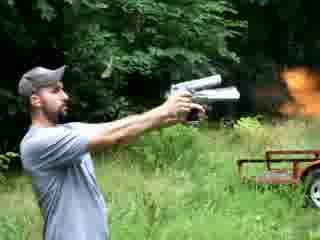}
   	 	\includegraphics[width=0.12\textwidth]{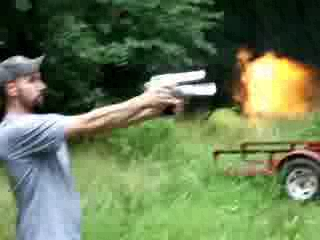}
   	    \includegraphics[width=0.12\textwidth]{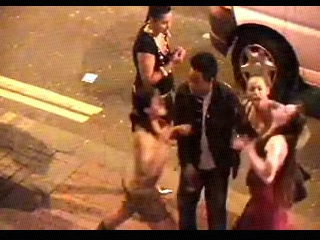}
   	 	\includegraphics[width=0.12\textwidth]{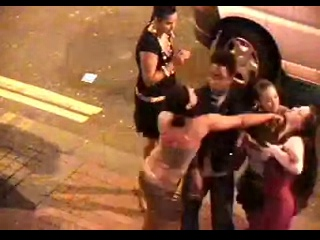}
   	 	\includegraphics[width=0.12\textwidth]{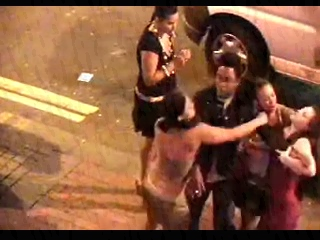}
   	 	\includegraphics[width=0.12\textwidth]{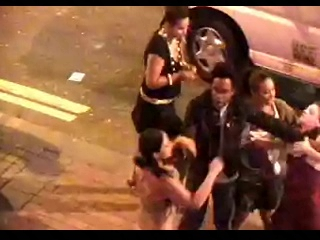}
   	 	}
   	 	\subfigure[C-DUP \cite{LiNPSKRS19}]{
   	 	\hspace{-0.1in}
	    \includegraphics[width=0.12\textwidth]{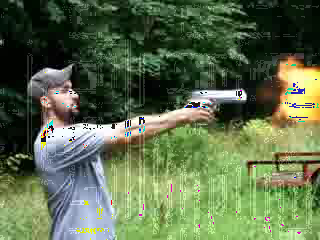}
  	    \includegraphics[width=0.12\textwidth]{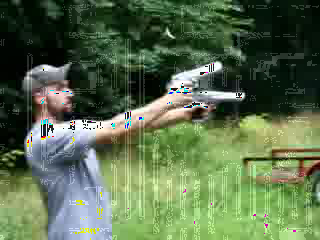}
   	 	\includegraphics[width=0.12\textwidth]{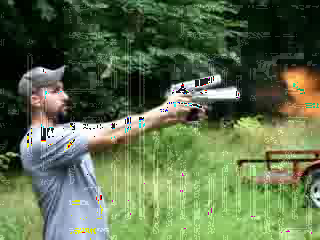}
   	 	\includegraphics[width=0.12\textwidth]{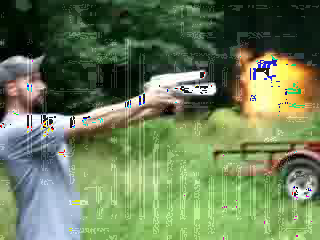}
   	    \includegraphics[width=0.12\textwidth]{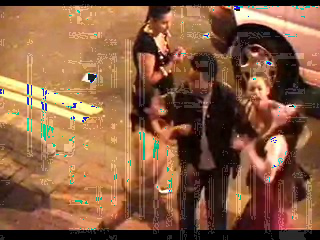}
   	 	\includegraphics[width=0.12\textwidth]{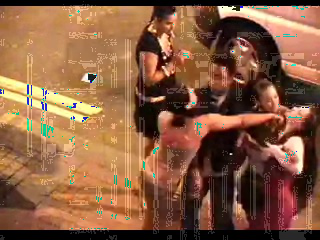}
   	 	\includegraphics[width=0.12\textwidth]{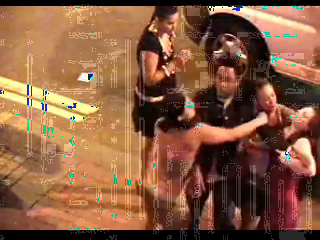}
   	 	\includegraphics[width=0.12\textwidth]{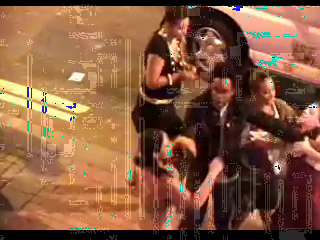}}

   	 \vspace{-0.1in}
   	 \caption{Some selected frames in videos corresponding to ``shooting'' and ``fighting'': (a) perturbed by U3D$_p$, (b) perturbed by U3D$_g$ and (c) perturbed by C-DUP \cite{LiNPSKRS19}. Both U3D$_p$ and U3D$_g$ show good human-imperceptibility compared with C-DUP.} 
	%\vspace{-0.1in}
	\label{fig:8}
\end{figure*}

\begin{table}[h]
 \centering\small
 \caption{Adversarial training on UCF Crime}
 
  \begin{tabular}{c|c|c|cc}
  \hline
  
 \multirow{2}{*}{Model}&\multirow{2}{*}{Defense} &{Clean}&{U3D$_p$} &{U3D$_g$}   \\
 && ACR&(SR)&(SR)\\
  \hline
  \multirow{3}{*}{C3D} & Normal& 92.5\% & 91.6\% & 90.7\%
  \\&UAT  & 84.5\% & 74.7\% & 75.3\% \\ 
    &U3D-AT  &  82.4\% & 62.7\% & 65.3\% \\ 
      \hline

  \multirow{3}{*}{I3D}&  Normal& 95.3\% & 88.4\%& 91.2\%\\
 
  & UAT  &   89.4\% & 76.2\% &80.5\%\\
    &U3D-AT   & 86.2\% & 58.6\% &59.5\%\\
  \hline
  
  \end{tabular}
  \label{tab:atcrime}
\end{table}

\begin{table}[h]
 \centering
 \caption{Randomized smoothing on UCF Crime} 
  \small
  \begin{tabular}{c|c|c|cc|cc}
  \hline
  
 \multirow{2}{*}{Model} &\multirow{2}{*}{n} &{Clean} & \multicolumn{2}{c|}{U3D$_p$} & \multicolumn{2}{c}{U3D$_g$}   \\ \cline{4-5}\cline{6-7}
   & &ACR &ACR & Radius    &  ACR & Radius  \\
  \hline
 
  \multirow{2}{*}{C3D}& 100&70.6\% &  26.3\% & 0.26 &25.1\% & 0.22 \\ 
    & 1000 & 71.2\%  &  30.2\% & 0.21 &32.4\% & 0.23 \\ 
      \hline

  \multirow{2}{*}{I3D}& 100& 74.6\% &  28.3\% & 0.23 &25.6\% & 0.20 \\ 
    & 1000 & 75.1\%  &  29.2\% & 0.25 &26.4\% & 0.18 \\
  \hline
  
  \end{tabular}
  \label{tab:certified2}
\end{table}

\begin{table}[htbp]
 \centering\small
 \caption{Amortized runtime (each frame) for attacking the streaming video on UCF Crime (in Seconds)}  
  \begin{tabular}{c|cc|c}
  \hline
  
   Video Name   &  Codec & Inject  & Runtime  \\
  \hline
  
  Arson   & 0.010 &0.004   & 0.014\\
  Assault   & 0.010 &0.005        & 0.015 \\
  Explosion   &0.009 & 0.007  & 0.016      \\
  Fighting & 0.009 &  0.006        &  0.015\\
  Shooting   & 0.010 &0.004  & 0.014       \\
  \hline
  
  \end{tabular}
  \label{tab:time2}
\end{table}

Second, Table \ref{tab:atcrime} and \ref{tab:certified2} present the additional results for U3D against defense schemes. All the results have confirmed the high attack performance even against defense schemes (similar to the results in Section \ref{exp:defend}). Table \ref{tab:time2} presents the amortized runtime (each frame) for attacking the streaming video (i.e., five videos in the UCF Crime).

\subsection{Particle Swarm Optimization for U3D Parameters}
\label{sec:pso}

PSO was first developed for social behavior simulation \cite{eberhart1995new,clerc2002particle,hassan2005comparison, van2007analysis} and then commonly applied to optimization, e.g., model selection \cite{escalante2009particle}. Moreover, PSO can be approximated for the optimization with a high-dimensional intense search space and numerous local optimal. 

To utilize PSO to optimize U3D perturbations, we first define the fitness function $\mathcal{A}(f,V,\mathcal{N}(T;\vec{s_i}))$, detailed in Algorithm \ref{alg:oracle}. Then we aim to find the optimal particle position (i.e., U3D parameters value) for such fitness function. Algorithm \ref{alg:swarm} demonstrates the detailed process of U3D optimization. 
At the beginning, a swarm of $m$ particles denoted as $S=\{\vec{x_1}^k, \vec{x_2}^k, \dots, \vec{x_m}^k\}$ will be initialized. 
For each iteration $k$, each particle $i$ holds a position $\vec{x_i}^k=[x_{i,1}^k, x_{i,2}^k, \dots, x_{i,d}^k]$, where $d$ is the dimension of the searching parameter space and $x_{i, j}, j\in [1, d]$ indicates the parameter value of $j$th dimension. To update its position $\vec{x_i}^{k+1}=\vec{x_i}^{k}+\vec{v_i}^k$, each particle $i$ compute with its current velocity $\vec{v_i}^k=[v_{i,1}^k, v_{i,2}^k, \dots, v_{i,d}^k]$ as the following equations: 
\vspace{-0.01in}
\begin{equation}
    v_{i,j}^{k+1}=W*v_{i,j}^k+c_1*r_1(s_{i,j}-x_{i,j}^k)+c_2*r_2(s_{g,j}-x_{i,j}^k) \label{eq:pso_v}
\end{equation}
\begin{equation}
    x_{i,j}^{k+1}=v_{i,j}^{k}+x_{i,j}^k \label{eq:pso_x}
\end{equation}

where (1) $s_{i,j}$ is the value for $k$th dimension of the best solution searched via particle $i$ so far; $S_i=[s_{i,j}], j\in[1, d]$ is called personal best; (2) $s_{g,j}$ is the value for $k$th dimension of the best solution in the Swarm $S$ so far; $S_g=[s_{g,j}], j\in[1, d]$ is called leader. Note that every particle can use the $S_i$ (local information) and $S_g$ (social information) to iteratively update its velocity and position. $c_1, c_2 \in \mathbf{R}$ are weights for quantifying the impacts of the personal and social best solution correspondingly; $r_1, r_2$ is uniformly distributed values of range $[0,1]$ which represents of randomness in the search. $W=\{W_s, W_f, W_e\}$ is called inertia weight, which can control the impacts of the previous velocity on the current iteration, and then influence searching ability. $W$ will be decreased with every iteration via the following equation: $W=W-\frac{W_s-W_e}{k*W_f}$, where $W$ is initialized as $W_s$ and ended as $W_e$.

\begin{algorithm}
\small
\SetKwComment{Comment}{ $\triangleright$\ }{}

\KwIn{public DNN model $f$, public video dataset $V$, current U3D noise parameters $\vec{s}$, sample times $I$}
\KwOut{Output value $r$ of fitness function}
Initialize  $r\gets 0$

$\xi \gets \mathcal{N}_p$

\For{$v_i\in V$}
{
    
    $t\gets 0$  
    
    \tcp{Sample I times}
    
    \For{$i\in I$}
    {
    $\tau\gets U[0, T-1]$
    
    $t\gets t+\mathcal{D}(v_i,v_i+\mathtt{Trans}(\xi, \tau);d)$
    }
    
    $r\gets r+\frac{t}{I}$

}

$r\gets \frac{r}{|V|}$

\Return $r$

\caption{Attack Objective $\mathcal{A}(f, V,  \mathcal{N}(T;\vec{s}))$}\label{alg:oracle}
\end{algorithm}

\begin{algorithm}
\small
    \caption{$\mathtt{NoiseOpt}$($f$, $V$)}
    \label{alg:swarm}
    \KwIn{U3D function $\mathcal{N}(\cdot)$, DNN model $f$,  video dataset $V$, $\ell_{\infty}$-norm bound $\epsilon$, search space $\mathcal{X}$ for U3D perturbation parameter $\mathcal{S}$; PSO model: inertia weight $W=\{W_s, W_f, W_e\}$, individual/social weights $c_1, c_2$, swarm size $m$, maximum iteration number $h$}
    \KwOut{optimal parameter set $\mathcal{S}^*$}

    \tcp{each node has $||(\mathcal{N}(T;\vec{s_i}))||_\infty \leq \epsilon$}
    $\mathcal{X}_{sample} \gets$ randomly sample $m$ points from $\mathcal{X}$
    
    \For{each $\vec{s_i} \in \mathcal{X}_{sample}$}
    {
    
    Call Algorithm \ref{alg:oracle}: $\mathcal{A}(f,V,\mathcal{N}(T;\vec{s_i}))$
    
    Set personal best of each particle $S_i\gets \vec{s_i}$}
    
    Find the leader $S_{gb}$

    Initialize $\vec{s_i}^k \gets \vec{s_i}, i\in[1, m]$
    
    \While{$k=1\leq h$}{
    \For{$i\in[1, m]$}{

    Update velocity and position per Equation \ref{eq:pso_v}, \ref{eq:pso_x}
    
    Repeat Line 2-4

    }
    
    Update $S_{gb}$ if leader changes
    
    Update inertia weight $W$
   
    }
    \Return $S_{gb}$ as $S^{*}$  
    
\end{algorithm}
\subsection{Comparison of PSO with Other Meta-Heuristic Algorithms}\label{app:pso_performance}

We use the C3D model as the public DNN model and randomly sample 500 videos from the HMDB51 dataset as the public dataset. The $\epsilon$ is set as a small bound 8. The parameter of the normalization $\alpha$ is set to 0.5. Table \ref{tab:para} shows the specified value ranges of the parameters for U3D$_p$ and U3D$_g$, respectively. As for PSO, we set up the parameters as follows: (1) swarm size $m=20$; (2) individual and social weight $c_1=c_2=2$; (3) inertia weight $W=\{1.2, 0.5, 0.4\}$; (4) maximum iteration times $h=40$. 

\begin{table}[htbp]
 \center\small
 \caption{U3D parameters setting}
  \begin{tabular}{ccc|ccc}
  \hline
  
&U3D$_p$ & & &U3D$_g$\\
\hline
  $\lambda_x, \lambda_y, \lambda_t$    &  $\Lambda$  & $\phi$   & $K$ & $\sigma$    &   $F$ \\
  \hline
  
$[2, 180]$   &  $[1, 5]$  &   $[1, 60]$   &$[1,5]$ &  $[1, 20]$  &   $[0.25, 20]$ \\
 
  \hline
  \end{tabular}
  \label{tab:para}
 
\end{table}

Then, we compare PSO with genetic algorithm (GA) \cite{hassan2005comparison}, simulated annealing (SA) \cite{van1987simulated}, and Tabu search (TS) \cite{glover1989tabu} on tuning the U3D parameters. For GA, we set the number of chromosomes to be $20$ (same as the number of PSO's particles) with the combination of tournament selection with a 50\% uniform crossover probability \cite{hassan2005comparison} and mutation rate 0.5\%. For SA, we set the initial temperature is 5000, and cooling factor 0.99. For TS, the tabu list size is set to 4. We implement the four methods for both U3D$_p$ and U3D$_g$ on the 500 videos, which are repeated 5 times and averaged for the final results. Table \ref{tab:optcom} illustrates their experimental results. We can observe that the PSO-based method is efficient, and also slightly outperforms GA, SA and TS on attack performance, e.g., PSO improves 1.7\% over GA and 3.3\% over SA for U3D$_p$. Besides, the MSE of both U3D perturbations are below 20 (very minor distortion out of 255$^2$ in the scale).

\begin{table}[htbp]
 \centering\small
 \caption{PSO vs. GA, SA and TS (learning U3D parameters offline) for U3D$_p$ and U3D$_g$ (success rate ``SR'').} 
  \begin{tabular}{c|ccc|ccc}
  \hline

   \multirow{2}{*}{Method} &   \multicolumn{3}{c|}{U3D$_p$}  &  \multicolumn{3}{c}{U3D$_g$} \\ \cline{2-4} \cline{5-7} & Time (s)  & SR &MSE & Time (s) & SR& MSE\\
  \hline
  
PSO    &   847  &  88.7\%      & 15.3 & 789&89.6\%& 16.0\\
  GA  &  1,481  &  87.0\%        &  14.6 &1,164  &  89.4\%&17.8\\
   SA  &  1,976   &     85.4\%     &16.9 &  2,267 & 87.7\% &13.7\\
     TS  & 1,039 &86.5\%        & 17.5   & 822 &  88.1\%&15.8\\
  \hline
 
  \end{tabular}
  \label{tab:optcom}
\end{table}

\begin{figure*}[!tbh]
\center
	\includegraphics[angle=0, width=1\linewidth]{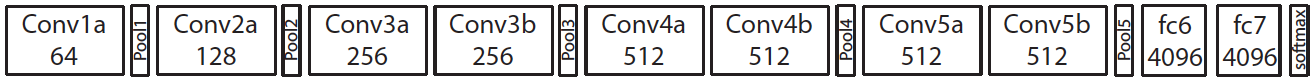}
	\caption[Optional caption for list of figures]
	{The C3D architecture \cite{TranBFTP15} consists of 8 convolution, 5 max-pooling, and 2 fully connected layers, followed by a softmax output layer. All 3D convolution kernels are 3 $\times$3 $\times$3 with a stride of 1 in both spatial and temporal dimensions. The number of filters is denoted in each box \cite{TranBFTP15}. The 3D pooling layers are represented as pool1 to pool5. All pooling kernels are 2 $\times$ 2 $\times$ 2, except for pool1, which is 1 $\times$ 2 $\times$ 2. Each fully connected layer has 4,096 output units.}
	\label{fig:c3d}
\end{figure*}

\subsection{Discussions}
\label{sec:add_dis}

\noindent\textbf{U3D Stealthiness}. Stealthiness of adversarial attacks can be reflected with two properties: (1) human-imperceptibility with minor perturbations; (2) misclassified outputs are not always rare. U3D can ensure much better human-imperceptibility (as validated in Section \ref{sec:exp}), compared to the white-box attack C-DUP \cite{LiNPSKRS19}. For the latter one, as a black-box attack, adversaries of U3D do not know the class labels and their distributions (different from the white-box attack \cite{LiNPSKRS19}), and inject the U3D perturbations to misclassify the perturbed videos to ``random class labels''. If the adversaries prefer to further improve stealthiness by ``supervising'' such misclassification to ``common class labels'' in the attack, they can query the target DNN model (similar to the traditional black-box attack) to retrieve some class labels and their distribution, and then adjust the objective function/constraint(s) in the U3D parameters learning to tune the U3D perturbation towards such goal. This also applies to other DNN-based video recognition systems, such as anomaly detection. We will explore these in the future. 

\vspace{0.05in}

\noindent\textbf{Potential Advanced Attacks}. Our U3D obtains good transferability to attack the video recognition systems in the black-box setting. To further improve the attack performance, we can utilize the query-based black-box attack to determine the attack gradient direction with a limited number of queries \cite{chen2017zoo, chen2020hopskipjumpattack}. We can also improve the U3D attack via learning the U3D perturbations on an ensemble of diverse models \cite{tramer2017ensemble} or diverse inputs \cite{xie2019improving}. In summary, the adversarial attack can be more adaptive in the arms race.

\vspace{0.05in}

\noindent\textbf{Attacking New and Future DNNs}. To date, besides the commonly-used C3D \cite{TranBFTP15} and I3D \cite{CarreiraZ17}, new DNN models built on the 3D spatio-temporal convolutions are proposed all the time, such as R(2+1)D \cite{TranWTRLP18}, and CSN \cite{abs-1904-02811}. For example, the recently proposed Channel Separated Convolutional Network (CSN) \cite{abs-1904-02811} factorizes 3D convolution via separating spatio-temporal and channel interactions, which improves both the accuracy and computation efficiency. Considering that our U3D perturbations work directly on spatio-temporal features across consecutive frames, we anticipate that those new spatio-temporal features may also be vulnerable to U3D perturbations. We plan to experimentally validate such vulnerabilities (in case of U3D perturbations) in the newly proposed and future DNN models as soon as they have been deployed.

\vspace{0.05in}

\noindent\textbf{Attacking Image Classifiers}. It is straightforward to downgrade our U3D attack to DNN models on images. Since the image can be considered as a 1-frame video. We only need to fix $t$ to 1 to generate the 2-D (frame) perturbation.

\vspace{0.05in}

\noindent\textbf{Physical U3D Attack}. The U3D attack can be extended to the physical adversarial attack in real world \cite{kurakin2016adversarial}. For example, our U3D perturbations can be integrated with the visual light technology (e.g., smart LED) \cite{vla19,zhou2018invisible}, with which the U3D perturbations can be projected (realizing the U3D perturbations with more programmable light building block) on the scenes inconspicuously. In addition, with computer graphic primitives, our U3D perturbations look visually like natural textures, then the manipulated light will not easily be discerned by humans. We will explore this in the future.

\subsection{Video Recognition Systems and Models}
\label{sec:vrs}

\subsubsection{Summary of Difference among Video DNN Models} \label{app:difference}

The five DNN models (C3D, I3D, DN, LRCN, and TSN) in the transferability evaluations are very different. 

First, C3D and I3D are using 3D convolution (3DConv) kernels, while DN, LRCN, and TSN are using 2D convolution kernels. Therefore, C3D and I3D can naturally capture temporal information in their convolution operations, while DN, LRCN, and TSN require additional components to capture the temporal information. Second, C3D is a shallow network with only 8 convolution layers, and its convolution kernel sizes are predefined. Although I3D is also based on the 3D (spatio-temporal) feature extraction, it requires another extra stream of optical flow information, which is also very different from the C3D in the model architecture. Furthermore, other three models (DN, LRCN, and TSN) have totally different architectures. DN is purely based on 2D convolution networks for images. It uses 2D convolution to extract features from each video frame, and then fuses features from multiple frames. LRCN also uses a 2D convolution network, but it uses a long short-term memory (LSTM) to fuse features from multiple frames. TSN uses two convolution networks, one for video frames, and the other one for optical information computed from frames.

\subsubsection{C3D Network}
\label{app:c3d}

Figure \ref{fig:c3d} shows the detailed structure and the characteristics of the C3D network.

\subsubsection{DNN-based Anomaly Detection}
\label{app:anomaly}
\begin{figure}[!tbh]
\includegraphics[width=1\linewidth]{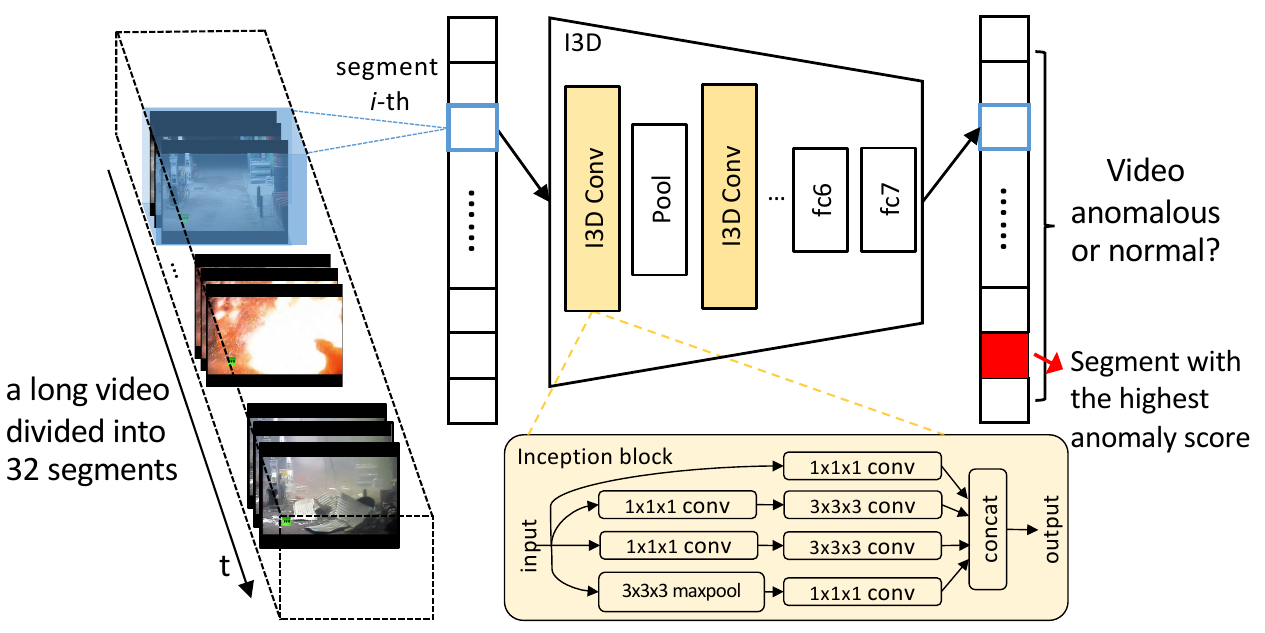}
\caption{Anomaly detection \cite{SultaniCS18} with I3D.}
\vspace{-0.05in}
\label{fig.anomaly_detection}
\end{figure}

\begin{figure}[!tbh]
\center
\includegraphics[width=0.66\linewidth]{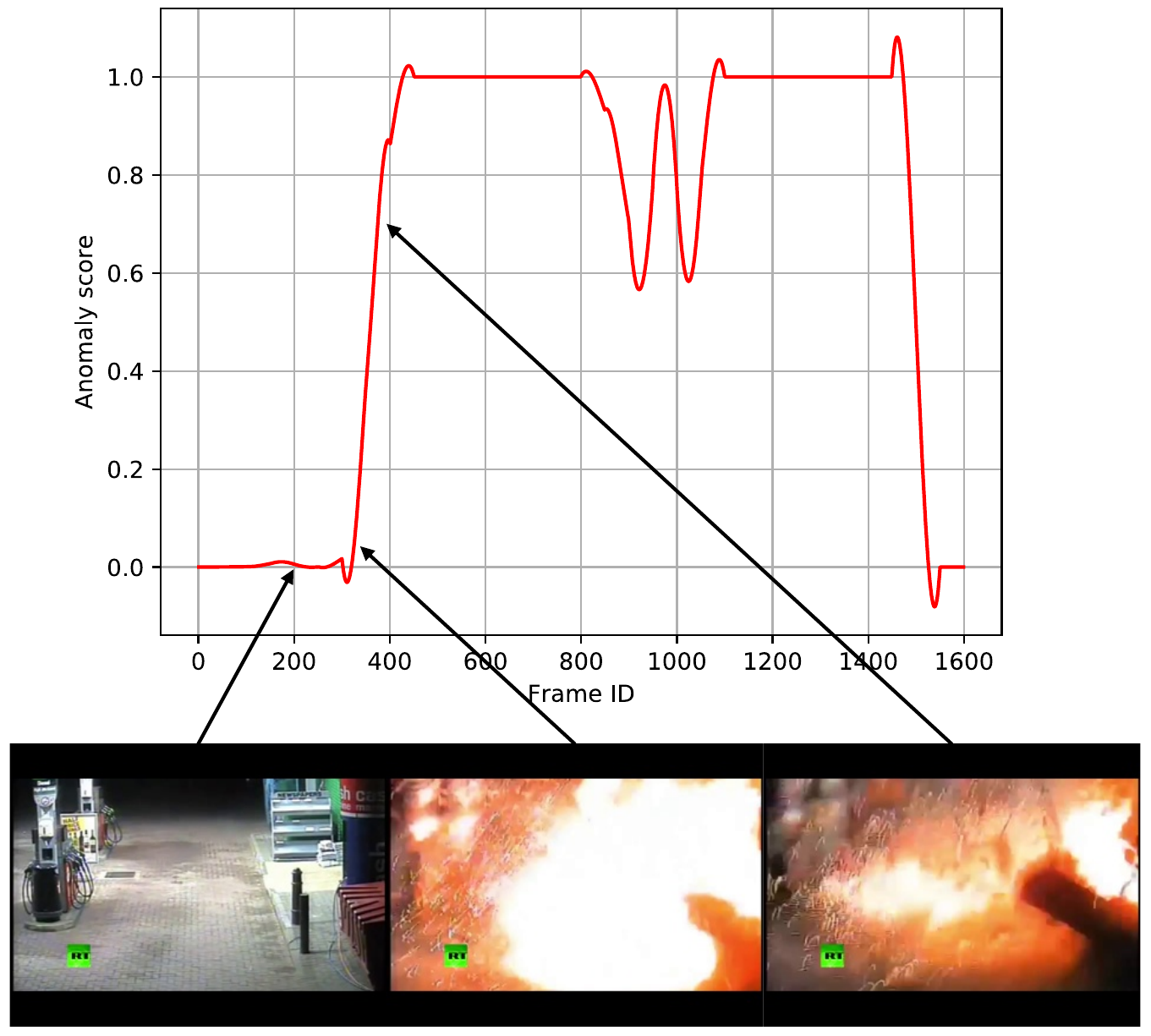}

\vspace{0.01in}
\caption{Video anomaly detection system for detecting an ``Explosion'' event. The score increases from 0 (normal scenery) to high (explosion).}

\label{fig.anomaly}
\end{figure}

Figure \ref{fig.anomaly_detection} illustrates a video anomaly detection system with the I3D model (by integrating additional optical flow information into the spatio-temporal features). Figure \ref{fig.anomaly} illustrates
the curve of anomaly scores inferred by the I3D model in time series. At first, the score is close to 0 for the normal scenery. Then, it increases as the explosion occurs to report such anomalous event (with a threshold).

\end{document}